\documentclass[prb,superscriptaddress,showpacs,floatfix,tightenlines,10pt,twocolumn]{revtex4}
\usepackage{amssymb}
\usepackage{amsmath}
\usepackage{graphicx}

\begin{document}

\title{Quantum Hall ferromagnetic phases in the Landau level $N=0$ of a
graphene bilayer}
\author{J. Lambert}
\affiliation{D\'{e}partement de physique, Universit\'{e} de Sherbrooke, Sherbrooke, Qu%
\'{e}bec, J1K 2R1, Canada}
\author{R. C\^{o}t\'{e} }
\affiliation{D\'{e}partement de physique, Universit\'{e} de Sherbrooke, Sherbrooke, Qu%
\'{e}bec, J1K 2R1, Canada}

\begin{abstract}
In a Bernal-stacked graphene bilayer, an electronic state in Landau level $%
N=0$ is described by its guiding-center index $X$ (in the Landau gauge) and
by its valley, spin, and orbital indices $\xi =\pm K,\sigma =\pm 1,$ and $%
n=0,1.$ When Coulomb interaction is taken into account, the chiral
two-dimensional electron gas (C2DEG) in this system can support a variety of
quantum Hall ferromagnetic (QHF)\ ground states where the spins and/or
valley pseudospins and/or orbital pseudospins collectively align in space.
In this work, we give a comprehensive account of the phase diagram of the
C2DEG at integer filling factors $\nu \in \left[ -3,3\right] $ in Landau
level $N=0$ when an electrical potential difference $\Delta _{B}$ between
the two layers is varied. We consider states with or without layer, spin, or
orbital coherence. For each phase, we discuss the behavior of the transport
gap as a function of $\Delta _{B},$ the spectrum of collective excitations
and the optical absorption due to orbital pseudospin-wave modes. We also
study the effect of an external in-plane electric field on a coherent state
that has both valley and spin coherence and show that it is possible, in
such a state, to control the spin polarization by varying the strength of
the external in-plane electric field.
\end{abstract}

\date{\today }
\maketitle

\section{INTRODUCTION}

Electrons in a Bernal-stacked graphene bilayer\cite%
{Bilayerreview,Castrojphys} behave as a chiral two-dimensional gas of
massive Dirac fermions (C2DEG)\cite{Barlas}. The chiral nature of the
electrons lead to transport and optical properties that are different from
those of conventional semiconductor 2DEG'S or of the 2DEG in monolayer
graphene. In particular, in the absence of Coulomb interaction, the Landau
level (LL)\ spectrum is given by $E_{N}=\pm \hslash \omega _{c}^{\ast }\sqrt{%
N\left( N+1\right) },$ where the Landau level index $N=0,\pm 1,\pm 2,...$
and the effective cyclotron frequency $\omega _{c}^{\ast }=eB/m^{\ast }c$
where $B$ is the magnetic field and $m^{\ast }$ the effective mass of the
electrons. Each Landau level is four-time degenerate when counting valley
and spin degrees of freedom with the exception of Landau level $N=0$ which
is eight-time degenerate. Indeed, an electronic state in Landau level $N=0$
is specified by its guiding-center $X$ (in the Landau gauge), spin $\sigma
=\pm 1,$ valley $\xi =\pm K$ and orbital $n=0,1$ indices\cite{Bilayerreview}%
. (In Landau level $N=0,$ valley and layer degrees of freedom are
equivalent.) When the Coulomb interaction is negligible with respect to the
disorder broadening at low temperature and when the small Zeeman splitting
is neglected, the eight states in $N=0$ are degenerate and the Hall
conductivity has plateaus at $\sigma _{xy}=4Me^{2}/h$ where $M=\pm 1,\pm
2,...$\cite{Novoselov}.

In recent transport experiments\cite%
{Feldman,Martin,Bao,Dean,Zhao,Velasco,Freitag,Weitz}, it was shown that in
sufficiently pure sample, when disorder is low or when the magnetic field is
large enough, the Coulomb interaction completely lifts the degeneracy of the 
$N=0$ octet and lead to the formation of seven new plateaus in the Hall
conductivity i.e. $\sigma _{xy}=\nu Me^{2}/h,$ where $\nu \in \left[ -3,3%
\right] .$ These plateaus were attributed to the formation of
broken-symmetry many-body ground states. These states can alternatively be
described as quantum Hall ferromagnets (QHF's) where the spin and/or valley
pseudospins and/or orbital pseudospins are spontaneously and collectively
aligned in space\cite{BarlasPRL1}.

In bilayer graphene, a top-bottom gates voltage imbalance can be applied to
create a potential difference $\Delta _{B}$ (which we call the "bias"
hereafter) between the two layers. In dual-gated bilayer graphene, $\Delta
_{B}$ and the total density of electrons in the bilayer can be controlled
independently. This allows the phase diagram of the C2DEG to be studied as a
function of $\Delta _{B},$ magnetic field and temperature. Such study has
been done by Weitz et al.\cite{Weitz} in high-quality bilayer graphene
suspended between a top gate electrode and the substrate. The measurements
show a series of phase transitions between different QHF states as $\Delta
_{B}$ is increased at given filling factor and magnetic field. Special
attention has been given to the filling factor $\nu =0$ where the precise
nature of the ground state when $B\rightarrow 0$ near zero bias is still
debated\cite{Kim}. All experiments were done at relatively small magnetic
field $B<10$ T with the exception of the experiments reported in Ref. %
\onlinecite{Zhao} where $B$ reached $35$ T.

Various aspects of the QHF\ states in bilayer graphene (in particular the
nature and the evolution of the ground state of the C2DEG near charge
neutrality as $B\rightarrow 0$) have been studied theoretically by a number
of authors\cite%
{Nandkishore,Gorbarnu0,Gorbarnutot,Gorbarconductivity,Shizuya2012,Shizuyadipole,Shizuyacyclotron,Shizuyamodes01,Sari}%
. In the work of Gorbar et al.\cite{Gorbarnutot} and Shizuya\cite%
{Shizuya2012}, the phase diagram of the C2DEG as a function of bias for $\nu
=0,1,2,3$ is presented. Gorbar et al. have considered the effect of both
static\cite{Gorbarnutot} and dynamical screenings\cite{Gorbarconductivity}
of the Coulomb interaction. The modified gap equation captures the linear
scaling of the transport gaps with the magnetic field which is seen in all
the experiments at low magnetic field. The transport gaps are also strongly
reduced with respect to the unscreened case and become more comparable to
those observed experimentally. The dynamical screening was found to
reproduce the offset in the behavior of the gap with magnetic field in the
spin-polarized QHF\ state near zero bias which is seen in the experiments%
\cite{Velasco}. Shizuya points out that the filled levels $N\leq -1$ cannot
be considered as completely inert. Instead, they lead to a correction $%
\Lambda _{n}$ of the energy of the orbital levels $n=0,1$ that can change
the ordering of these states in a way that depend on their occupation. A
negative capacitance effect is also found that suppress rotation of the
valley pseudospins.

The work that we present in this paper extends our previous study of the
phase diagram of the C2DEG at zero bias\cite%
{BarlasPRL1,BarlasPRL2,CoteOrbital,CoteOrbital2} and complements the study
of Gorbar et al. and Shizuya. We give a comprehensive account of the phase
diagram of the C2DEG in bilayer graphene at all integer filling factors $\nu
\in \left[ -3,3\right] $ in Landau level $N=0$ as a function of an applied
bias $\Delta _{B}$ at a fixed magnetic field. Our analysis is based on an
effective two-band model\cite{Mccann} which describes the low-energy physics
near the valleys $K_{\pm }$. We explicitly take into account as
symmetry-breaking terms the\ Zeeman splitting $\Delta _{Z}$ and the bias $%
\Delta _{B}$. In the effective two-band model, the bias $\Delta _{B}$ lifts
the degeneracy between the orbital levels $n=0$ and $n=1$ by a small amount $%
\xi \beta \Delta _{B}$ where $\beta =\gamma _{1}/\hslash \omega _{c}^{\ast }$
with $\gamma _{1}$ the interlayer hopping between carbon atoms that are part
of a dimer. We also include in our model the interlayer next
nearest-neighbor hopping term $\gamma _{4}$ between carbons atoms in the
same sublattices. This term causes a small asymmetry in the electronic band
structure and was neglected in previous studies\cite{Gorbarnutot,Shizuya2012}%
. It combines with the correction $\xi \beta \Delta _{B}$ to give an energy
difference $\approx 2\beta \gamma _{1}\gamma _{4}/\gamma _{0}+\xi \beta
\Delta _{B}$ between the $n=1$ and $n=0$ orbital levels where $\gamma _{0}$
is the intralayer hopping energy between nearest-neighbors. This correction
is thus finite at zero bias and breaks the orbital degeneracy. Our phase
diagram is not electron-hole symmetric around $\nu =0$ and the sequence of
phase transitions is different for each filling factor.

In our analysis, we treat the electron interaction in the\ Hartree-Fock
approximation (HFA) and compute the collective excitations and
electromagnetic absorption of the different phases of the C2DEG in the
generalized random-phase approximation (GRPA). We include in our study both
uniform and non-uniform states and allow for the possibility of any type of
coherent (or QHF) state. By coherent state, we mean a state where the
average value $\left\langle c_{\xi ,\sigma ,n,X}^{\dag }c_{\xi ^{\prime
},\sigma ^{\prime },n^{\prime },X}\right\rangle \neq 0$ for $\xi \neq \xi
^{\prime }$ and/or $\sigma \neq \sigma ^{\prime }$ and/or $n\neq n^{\prime }$
where $c_{\xi ,\sigma ,n,X}^{\dag }$ creates an electron in a state with
quantum numbers $\xi ,\sigma ,n,X.$ In our phase diagram which is summarized
on Fig. 5, the layer-coherent states occur at very small bias because of the
small interlayer distance $d=0.34$ nm in bilayer graphene. As the bias is
increased, we find around a critical bias corresponding to the regions where
the Hall conductivity ceases to be quantized in the experiments\cite{Weitz}
a state with both layer and spin coherence. The orbital coherent states
occur at a much larger bias corresponding to the situation where level $n=1$
gets lower in energy than level $n=0$ in valley $K_{-}$ (see Fig. \ref%
{figniveaux}). In-between these coherent states are various incoherent
states, some of which have been studied before\cite{Gorbarnutot}.
Interestingly, we find that the application of an electric field in the
plane of the layers can produce a new state where all three coherences
(layer, orbital and spin) are present. In such a state, it is possible to
control the degree of \textit{spin} polarization by changing the strength of
the external in-plane \textit{electric} field.

We also present a study the properties of the different ground states in the
phase diagram. For all filling factors, we show how the transport gaps
evolve with bias. In most cases, this evolution is qualitatively similar to
that obtained with screening corrections\cite{Gorbarnutot}. We compute all
the intra-LL collective excitations in the various phases showing that all
coherent states but the orbital state are characterized by a
linearly-dispersing gapless (in the long-wavelength limit) Goldstone mode.
This mode becomes gapped after the transition to an adjacent incoherent
state. In the orbital phase, the orbital-pseudospin Goldstone mode
dispersion is anisotropic and this mode becomes unstable at a finite wave
vector indicating a transition to a charge-density-wave state\cite%
{CoteOrbital,CoteOrbital2}. We identify the number of spin-waves and orbital
modes in each phase. These later modes are active in optical absorption. The
inter-LL and some intra-LL magnetoexcitons have been computed recently\cite%
{Sari,Shizuyadipole,Shizuyacyclotron,Shizuyamodes01} and we comment on the
difference with our results and how the presence of the inter-LL
magnetoexcitons in the spectrum may complicate the detection of the intra-LL
excitations. The main results of our paper are summarized in Fig. 5 (phase
diagram), Fig. 8 (transport gaps) and Fig. 9,10 (collective mode
dispersions).

This paper is organized in the following way. Section\ II introduces the
two-band model of bilayer graphene with the resulting LL spectrum in finite
magnetic field. Section III summarizes the Hartree-Fock and generalized
random-phase approximations that we use to take into account the Coulomb
interaction and gives the formalism for the calculation of the
electromagnetic absorption. Our numerical results for the phase diagram,
transport gaps, collective excitations and optical absorption are presented
in Sec. IV. In Sec. V, we show how the application of an in-plane electric
field allow to control the spin polarization in some phases. We conclude in
Sec. VI.

\section{TWO-BAND MODEL OF BILAYER GRAPHENE}

\subsection{Crystal structure and tight-binding Hamiltonian}

\begin{figure}[tbph]
\includegraphics[scale=0.8]{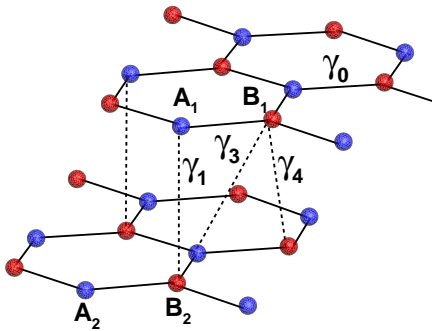}
\caption{(Color online) Crystal structure of a Bernal-stacked graphene
bilayer.}
\label{figstructurebicouche}
\end{figure}

The crystal structure of a Bernal-stacked graphene bilayer is shown in Fig. %
\ref{figstructurebicouche}. Each graphene layer is a two-dimensional crystal
with a honeycomb lattice structure. The honeycomb lattice can be described
as a triangular Bravais lattice with a basis of two carbon atoms $A_{n}$ and 
$B_{n}$ where $n=1,2$ is the layer index. The two basis vectors are given by 
$\mathbf{a}_{1}=a_{0}\left( 1/2,-\sqrt{3}/2\right) $and $\mathbf{a}%
_{2}=a_{0}\left( 1,0\right) ,$ where $a_{0}=2.\,\allowbreak 46$ \AA $=\sqrt{3%
}c$\ is the lattice constant of the underlying triangular Bravais lattice
and $c=1.42$ \AA\ is the separation between two adjacent carbon atoms. The
distance between the two graphene layers is $d=3.4$ \AA . In the Bernal
stacking arrangement, the upper $A$ sublattice is directly on top of the
lower $B$ sublattice while the upper $B$ sublattice is above the center of a
hexagonal plaquette of the lower layer.

The Brillouin zone of the reciprocal Bravais lattice is shown in Fig. \ref%
{figbzone}. We choose the two nonequivalent valley points to be%
\begin{equation}
K_{\xi }=\left( \frac{2\pi }{a_{0}}\right) \left( \xi \frac{2}{3},0\right) ,
\end{equation}%
where $\xi =\pm $ is the valley index.

\begin{figure}[tbph]
\includegraphics[scale=0.8]{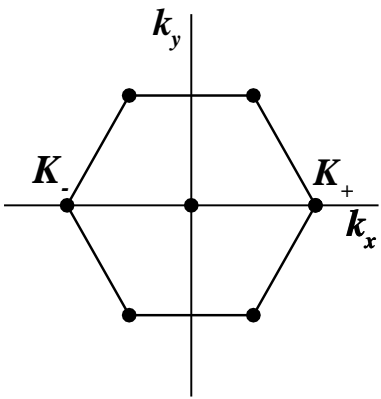}
\caption{Brillouin zone of the triangular Bravais lattice and definition of
the valleys indices $K_{\pm }.$}
\label{figbzone}
\end{figure}

The electronic dispersion is obtained from a tight-binding model with the
parameters\cite{Bilayerreview}:\ $\gamma _{0}$ the nearest-neighbor (NN)
hopping in each layer, $\gamma _{1}$ the interlayer hopping between carbon
atoms that are part of a dimer (i.e. $A_{1}-B_{2},$ these sites are called
the high-energy sites), $\gamma _{3}$ the interlayer NN hopping term between
carbon atoms of different sublattices (i.e. $A_{2}-B_{1}$) and $\gamma _{4}$
the interlayer next NN hopping term between carbons atoms in the same
sublattice (i.e. $A_{1}-A_{2}$ and $B_{1}-B_{2}$). The energy $\delta $
represents the difference in the crystal field between sites $A_{1},B_{2}$
and sites $A_{2},B_{1}$. In this work, we neglect the trigonal warping term $%
\gamma _{3}$, a correct approximation at sufficiently high magnetic field%
\cite{Mccann}.

If we define the spinor%
\begin{equation}
d_{\mathbf{k,}\sigma }^{\dag }=\left( 
\begin{array}{cccc}
a_{1,\mathbf{k,}\sigma }^{\dag } & b_{1,\mathbf{k,}\sigma }^{\dag } & a_{2,%
\mathbf{k,}\sigma }^{\dag } & b_{2,\mathbf{k,}\sigma }^{\dag }%
\end{array}%
\right) ,
\end{equation}%
where $a_{i,\mathbf{k,}\sigma }^{\dag }\left( b_{i,\mathbf{k,}\sigma }^{\dag
}\right) $ creates an electron on site $A(B)$ in layer $i=1,2$ with wave
vector $\mathbf{k}$ and spin $\sigma =\pm 1,$ then we can write the
second-quantized tight-binding Hamiltonian in the basis $%
(A_{1},B_{1},A_{2},B_{2})$ as

\begin{equation}
H^{0}=\sum_{\mathbf{k},\sigma }d_{\mathbf{k,}\sigma }^{\dag }H_{\sigma
}^{0}\left( \mathbf{k}\right) d_{\mathbf{k},\sigma },
\end{equation}%
with the matrix 
\begin{eqnarray}
&&H_{\sigma }^{0}\left( \mathbf{k}\right) =  \label{hamilcomplete} \\
&&\left( 
\begin{array}{cccc}
\begin{array}{c}
\frac{1}{2}\Delta _{B}+\delta \\ 
-\frac{1}{2}\sigma \Delta _{Z}%
\end{array}
& -\gamma _{0}\Lambda \left( \mathbf{k}\right) & -\gamma _{4}\Lambda ^{\ast
}\left( \mathbf{k}\right) & -\gamma _{1} \\ 
-\gamma _{0}\Lambda ^{\ast }\left( \mathbf{k}\right) & 
\begin{array}{c}
\frac{1}{2}\Delta _{B} \\ 
-\frac{1}{2}\sigma \Delta _{Z}%
\end{array}
& 0 & -\gamma _{4}\Lambda ^{\ast }\left( \mathbf{k}\right) \\ 
-\gamma _{4}\Lambda \left( \mathbf{k}\right) & 0 & 
\begin{array}{c}
-\frac{1}{2}\Delta _{B} \\ 
-\frac{1}{2}\sigma \Delta _{Z}%
\end{array}
& -\gamma _{0}\Lambda \left( \mathbf{k}\right) \\ 
-\gamma _{1} & -\gamma _{4}\Lambda \left( \mathbf{k}\right) & -\gamma
_{0}\Lambda ^{\ast }\left( \mathbf{k}\right) & 
\begin{array}{c}
-\frac{1}{2}\Delta _{B}+\delta \\ 
-\frac{1}{2}\sigma \Delta _{Z}%
\end{array}%
\end{array}%
\right) ,  \notag
\end{eqnarray}%
where $\Delta _{Z}=g\mu _{B}B$ with $g=2$ is the Zeeman energy. We have also
included in $H_{\sigma }^{0}$ an external transverse electric field that
creates an electrical potential difference (or bias) $\Delta _{B}$ between
the layers. The function $\Lambda \left( \mathbf{k}\right) $ is defined by%
\begin{equation}
\Lambda \left( \mathbf{k}\right) =\sum\limits_{i=1}^{3}e^{i\mathbf{k}\cdot 
\mathbf{\delta }_{i}},  \label{bg1}
\end{equation}%
where the summation is over the vectors connecting a site $A_{1}$ to its
three nearest-neighbors in the same plane i.e. $\mathbf{\delta }%
_{1}=a_{0}\left( 1/2,1/2\sqrt{3}\right) ,\mathbf{\delta }_{2}=a_{0}\left(
-1/2,1/2\sqrt{3}\right) ,\mathbf{\delta }_{3}=a_{0}\left( 0,-1/\sqrt{3}%
\right) .$

Using $\Lambda \left( \mathbf{K}_{\xi }+\mathbf{k}\right) \approx -\sqrt{3}%
a_{0}\left( \xi k_{x}+ik_{y}\right) /2$ and setting $\gamma _{4}=\delta
=\Delta _{B}=\Delta _{Z}=0,$ we find for the the band structure near the
points $K_{\xi }$ the bands 
\begin{eqnarray}
E_{1,\pm }\left( \mathbf{p}\right) &=&\pm \gamma _{1}\pm \frac{p^{2}}{%
2m^{\ast }}, \\
E_{2,\pm }\left( \mathbf{p}\right) &=&\pm \frac{p^{2}}{2m^{\ast }},
\end{eqnarray}%
where the momentum $\mathbf{p}$ is measured with respect to $\hslash \mathbf{%
K}_{\xi }$ and the effective mass is defined by%
\begin{equation}
m^{\ast }=\frac{2\hslash ^{2}\gamma _{1}}{3\gamma _{0}^{2}a_{0}^{2}}.
\end{equation}%
The band structure consists of four bands. In the absence of bias, the two
middle bands meet at the six valley points. The two high-energy bands are
separated by a gap $\gamma _{1}$ from the two middle, low-energy bands. The
bands $E_{2,\pm }\left( \mathbf{p}\right) $ remain degenerate at $\mathbf{p}%
=0$ when $\gamma _{3},\gamma _{4}$ and $\Delta $ are finite if $\Delta
_{B}=0.$ This degeneracy is lifted by a finite $\Delta _{B}$.

For a neutral bilayer, the chemical potential is at the energy $E=0.$ The
low-energy excitations ($E<<\gamma _{1}$) of the tight-binding model can be
studied using an effective two-band model\cite{Mccann,Macdotricouche}. This
model gives for each valley%
\begin{eqnarray}
&&H_{\xi ,\sigma }^{0}\left( \mathbf{p}\right) =  \label{hamilp} \\
&&\left( 
\begin{array}{cc}
\begin{array}{c}
\xi \frac{\Delta _{B}}{2}+\eta _{-\xi }p_{-}p_{+} \\ 
-\frac{1}{2}\sigma \Delta _{Z}%
\end{array}
& \frac{1}{2m^{\ast }}\left( p_{x}-ip_{y}\right) ^{2} \\ 
\frac{1}{2m^{\ast }}\left( p_{x}+ip_{y}\right) ^{2} & 
\begin{array}{c}
-\xi \frac{\Delta _{B}}{2}+\eta _{\xi }p_{+}p_{-} \\ 
-\frac{1}{2}\sigma \Delta _{Z}%
\end{array}%
\end{array}%
\right) ,  \notag
\end{eqnarray}%
where we used the basis $\left( A_{2},B_{1}\right) $ for $K_{-}$ and $\left(
B_{1},A_{2}\right) $ for $K_{+}$ and defined $p_{\pm }=p_{x}\pm ip_{y}$ and 
\begin{equation}
\eta _{\xi }=\frac{1}{2m^{\ast }}\left( \xi \frac{\Delta _{B}}{\gamma _{1}}+2%
\frac{\gamma _{4}}{\gamma _{0}}+\frac{\delta }{\gamma _{1}}\right) .
\end{equation}

In this model, the presence of a quantizing perpendicular magnetic field is
accounted for by making the Peierls substitution $\mathbf{p}\rightarrow 
\mathbf{P}=\mathbf{p}+e\mathbf{A}/c$ (with $e>0$), where $\nabla \times 
\mathbf{A=B=}B\widehat{\mathbf{z}}.$ Defining the ladder operators $a=\left(
P_{x}-iP_{y}\right) \ell /\sqrt{2}\hslash $ and $a^{\dag }=\left(
P_{x}+iP_{y}\right) \ell /\sqrt{2}\hslash $ with the magnetic length $\ell =%
\sqrt{\hslash c/eB},$ we get%
\begin{equation}
H_{\xi ,\sigma }^{0}=\left( 
\begin{array}{cc}
\begin{array}{c}
\xi \frac{\Delta _{B}}{2}+\zeta _{1,-}aa^{\dag } \\ 
-\frac{1}{2}\sigma \Delta _{Z}%
\end{array}
& \zeta _{2}a^{2} \\ 
\zeta _{2}\left( a^{\dag }\right) ^{2} & 
\begin{array}{c}
-\xi \frac{\Delta _{B}}{2}+\zeta _{1,+}a^{\dag }a \\ 
-\frac{1}{2}\sigma \Delta _{Z}%
\end{array}%
\end{array}%
\right) ,  \label{s1}
\end{equation}%
where%
\begin{eqnarray}
\zeta _{1} &=&\beta \left( 2\frac{\gamma _{1}\gamma _{4}}{\gamma _{0}}%
+\delta \right) , \\
\zeta _{1,\pm } &=&\zeta _{1}\pm \xi \beta \Delta _{B}, \\
\zeta _{2} &=&\beta \gamma _{1}\left( 1+2\frac{\delta \gamma _{4}}{\gamma
_{0}\gamma _{1}}+\left( \frac{\gamma _{4}}{\gamma _{0}}\right) ^{2}\right) ,
\end{eqnarray}%
and%
\begin{equation}
\beta =\frac{\hslash \omega _{c}^{\ast }}{\gamma _{1}}.
\end{equation}%
The effective cyclotron frequency is $\omega _{c}^{\ast }=eB/m^{\ast }c.$ In
Eq. (\ref{s1}), the ladder operators are defined such that $a^{\dag }\varphi
_{n}\left( x\right) =i\sqrt{n+1}\varphi _{n+1}\left( x\right) $ and $%
a\varphi _{n}\left( x\right) =-i\sqrt{n}\varphi _{n-1}\left( x\right) $
where $\varphi _{n}\left( x\right) $ with $n=0,1,2,...$ are the
eigenfunctions of the one-dimensional harmonic oscillator.

For all calculations done in this paper, we choose\cite{Castrojphys} for the
value of the parameters%
\begin{eqnarray}
\gamma _{0} &=&3.1\text{ eV,}  \label{param1} \\
\gamma _{1} &=&0.39\text{ eV,}  \label{param2} \\
\gamma _{4} &=&0.12\text{ eV,}  \label{param3} \\
\delta &=&0.0156\text{ eV.}  \label{param4}
\end{eqnarray}%
We have checked that the band dispersion obtained with this choice of signs
for the hopping terms is consistent with that reported in the literature\cite%
{Partoens}. With the magnetic field in Tesla, we have 
\begin{eqnarray}
\beta &=&8.\,\allowbreak 86\times 10^{-3}B, \\
\zeta _{1} &=&0.4\,\allowbreak 04B\text{ meV},
\end{eqnarray}%
while 
\begin{eqnarray}
\hslash \omega _{c}^{\ast } &=&3.\,\allowbreak 46B\text{ meV,} \\
\Delta _{Z} &=&0.1158B\text{ meV,} \\
\alpha &=&\frac{e^{2}}{\kappa \ell }=11.25\sqrt{B}\text{ meV.}
\end{eqnarray}%
In the calculation of $\alpha ,$ we take $\kappa \approx 5$ for the
effective dielectric constant at the position of the graphene layers. At $%
B=10$ T,

\begin{eqnarray}
\zeta _{1}/\alpha &=&3.\,\allowbreak 591\,1\times 10^{-2}\sqrt{B}=0.114\,, \\
\Delta _{Z}/\alpha &=&1.\,\allowbreak 029\,1\times 10^{-2}\sqrt{B}=0.0325, \\
\hslash \omega _{c}^{\ast }/\alpha &=&0.307\,56\sqrt{B}=0.973.
\end{eqnarray}

\subsection{Landau levels and eigenstates of the non-interacting Hamiltonian}

When $\gamma _{4}=\delta =\Delta _{B}=\Delta _{Z}=0,$ the Landau level
spectrum of $H_{\xi ,\sigma }^{0}$ is given by%
\begin{equation}
E_{N}^{0}=\mathrm{sgn}\left( N\right) \sqrt{\left\vert N\right\vert \left(
\left\vert N\right\vert +1\right) }\hslash \omega _{c}^{\ast },
\end{equation}%
where $N=0,\pm 1,\pm 2,...$ is the Landau level index and sgn is the signum
function. The corresponding eigenvectors of a given spin are 
\begin{equation}
\frac{1}{\sqrt{2}}\left( 
\begin{array}{c}
h_{\left\vert N\right\vert -1,X}\left( \mathbf{r}\right) \\ 
-\mathrm{sgn}\left( N\right) h_{\left\vert N\right\vert +1,X}\left( \mathbf{r%
}\right)%
\end{array}%
\right)
\end{equation}%
for $N\neq 0$. We use the Landau gauge $\mathbf{A}=\left( 0,Bx,0\right) $
where the eigenstates are 
\begin{equation}
h_{n,X}\left( \mathbf{r}\right) =\frac{1}{\sqrt{L_{y}}}e^{-iXy/\ell
^{2}}\varphi _{n}\left( x-X\right)
\end{equation}%
with $X$ the guiding-center index. All Landau levels $N\neq 0$ are four-fold
degenerate including spin and valley degrees of freedom in addition to the
guiding-center degeneracy $N_{\varphi }=S/2\pi \ell ^{2}$ where $S$ is the
area of the 2DEG. The Landau level $N=0$ is an exception because there are
two degenerate spinors with zero energy which are given, in the basis $%
\left( A_{2},B_{1}\right) $ for $K_{-}$ and $\left( B_{1},A_{2}\right) $ for 
$K_{+}$, by 
\begin{equation}
\left( 
\begin{array}{c}
0 \\ 
h_{0,X}\left( \mathbf{r}\right)%
\end{array}%
\right) ,\left( 
\begin{array}{c}
0 \\ 
h_{1,X}\left( \mathbf{r}\right)%
\end{array}%
\right) .  \label{bg2}
\end{equation}%
It follows that $N=0$ is eight-fold degenerate. In this paper, we restrict
the Hilbert space to the Landau level $N=0$ and use the index $n=0,1$ to
refer to the two "orbitals" $\varphi _{n=0}\left( x\right) $ and $\varphi
_{n=1}\left( x\right) $. With finite values of $\gamma _{4}$, $\delta
,\Delta _{Z}$ or $\Delta _{B}$, the valley, spin, and orbital degeneracies
are lifted and the noninteracting energies become 
\begin{eqnarray}
\;E_{\xi ,\sigma ,n=0}^{0} &=&-\frac{1}{2}\xi \Delta _{B}-\frac{1}{2}\sigma
\Delta _{Z},  \label{s10} \\
\;E_{\xi ,\sigma ,n=1}^{0} &=&-\frac{1}{2}\xi \Delta _{B}-\frac{1}{2}\sigma
\Delta _{Z}+\xi \beta \Delta _{B}+\zeta _{1}.  \label{s11}
\end{eqnarray}%
The corresponding eigenspinors are still given by Eq. (\ref{bg2}). Note that
the structure of the sublattice spinors in Eqs. (\ref{bg2}) is such that
states from different valleys (which are localized on different layers) have
no overlap. For $N=0$, the layer index is thus equivalent to the valley
index.

\subsection{Limit of validity of the two-band model}

\begin{figure}[tbph]
\includegraphics[scale=1]{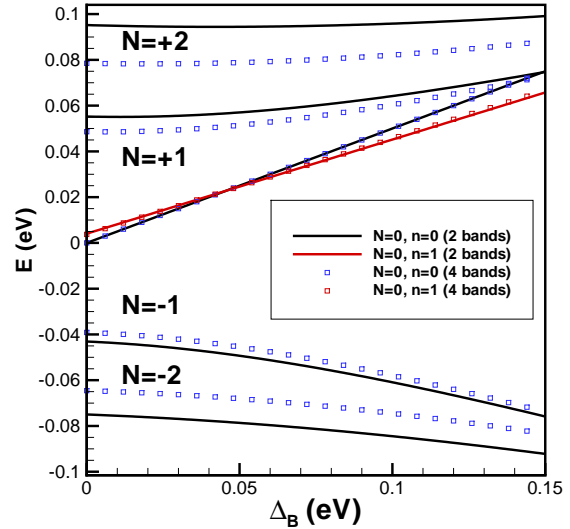}
\caption{(Color online) Comparison between the complete four-band model
(symbols) and the approximate two-band model (lines) for the electronic
dispersion in Landau levels $N=-2,-1,0,1,2$ and valley $K_{-}.$ }
\label{figcomparaisonmodeles}
\end{figure}

Fig. \ref{figcomparaisonmodeles} shows a comparison between the four-band
and two-band models for the electronic dispersion in Landau levels $%
N=-2,-1,0,1,2$ and valley $K_{-}$ using the values of the parameters given
previously. The agreement between the two models is excellent for $N=0$
where the difference in energy is of the order of $1\%.$ For levels $%
\left\vert N\right\vert >0,$ the difference in energy between the two models
is much more important. Note that the two sub-Landau levels of $N=0$
intersect level $N=1$ at $\Delta _{B}\approx 0.15$ eV. For the valley $K_{+}$
(not shown in the figure) the crossing occurs at a smaller bias $\Delta
_{B}\approx 0.10$ eV corresponding to an electric field $E_{\bot }\approx
300 $ meV/nm between the layers. In our calculation we must keep the bias
smaller than $\approx 0.10$ eV (i.e. $\Delta _{B}/\alpha \lesssim 2.8$ for $%
\kappa =5$) for our model to be valid.

\begin{figure}[tbph]
\includegraphics[scale=0.8]{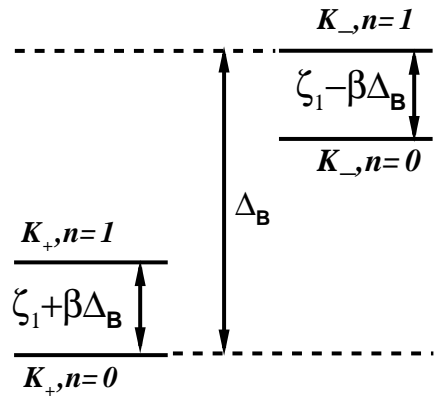}
\caption{Ordering of the four levels of a given spin in Landau level $N=0$
at finite bias $\Delta _{B}$. }
\label{figniveaux}
\end{figure}

Fig. \ref{figniveaux} shows the ordering of the four levels of a given spin
in $N=0$ at finite bias. The correction $\zeta _{1}$ opens a gap between the
two orbital states $n=0$ and $n=1$ which is independent of the bias. The
effective two-band model introduces a correction $\beta \Delta _{B}$ to this
gap that has different signs in the two valleys as indicated in the figure.
When combined with $\zeta _{1}$, the gap in valley $K_{+}$ is positive at
all biases while the gap in valley $K_{-}$ changes sign (level $n=1$ gets
below level $n=0$) when $\beta \Delta _{B}>\zeta _{1}$ i.e. for $\Delta
_{B}>0.046$ eV (i.e. $\Delta _{B}/\alpha =1.3$ for $\kappa =5$).

\section{INTERACTING CHIRAL TWO-DIMENSIONAL ELECTRON GAS}

We now add the Coulomb interaction to the noninteracting Hamiltonian.
Hereafter, we use the same basis $\left( A_{2},B_{1}\right) $ for both
valleys and define the field operators $\Psi _{\xi ,\sigma ,n}\left( \mathbf{%
r}\right) $ by 
\begin{equation}
\Psi _{-,\sigma ,n}\left( \mathbf{r}\right) =\sum_{X}\left( 
\begin{array}{c}
0 \\ 
h_{n,X}\left( \mathbf{r}\right)%
\end{array}%
\right) \otimes \left\vert \sigma \right\rangle c_{-,\sigma ,n,X},
\label{y1}
\end{equation}%
and%
\begin{equation}
\Psi _{+,\sigma ,n}\left( \mathbf{r}\right) =\sum_{X}\left( 
\begin{array}{c}
h_{n,X}\left( \mathbf{r}\right) \\ 
0%
\end{array}%
\right) \otimes \left\vert \sigma \right\rangle c_{+,\sigma ,n,X}.
\label{y2}
\end{equation}%
The second-quantized noninteracting part of the Hamiltonian is given by%
\begin{eqnarray}
H_{0} &=&\sum_{\sigma ,\xi ,n}\int d\mathbf{r}\Psi _{\xi ,\sigma ,n}^{\dag
}\left( \mathbf{r}\right) H_{\xi ,\sigma }^{0}\Psi _{\xi ,\sigma ,n}\left( 
\mathbf{r}\right) \\
&=&\sum_{\xi ,\sigma ,n}\sum_{X}\;E_{\xi ,\sigma ,n}^{0}c_{\xi ,\sigma
,n,X}^{\dag }c_{\xi ,\sigma ,n,X}.  \notag
\end{eqnarray}%
For the second-quantized Coulomb interaction, 
\begin{eqnarray}
V &=&\frac{1}{2}\sum_{n_{1},...,n_{4}}\sum_{\sigma ,\sigma ^{\prime
}}\sum_{\xi ,\xi ^{\prime }}\int d\mathbf{r}\int d\mathbf{r}^{\prime }\Psi
_{\xi ,\sigma ,n_{1}}^{\dag }\left( \mathbf{r}\right)  \label{vcoulomb} \\
&&\times \Psi _{\xi ^{\prime },\sigma ^{\prime },n_{2}}^{\dag }\left( 
\mathbf{r}^{\prime }\right) V_{\xi ,\xi ^{\prime }}\left( \mathbf{r}-\mathbf{%
r}^{\prime }\right) \Psi _{\xi ^{\prime },\sigma ^{\prime },n_{3}}\left( 
\mathbf{r}^{\prime }\right) \Psi _{\xi ,\sigma ,n_{4}}\left( \mathbf{r}%
\right) ,  \notag
\end{eqnarray}%
where the Coulomb potential 
\begin{equation}
V_{\xi ,\xi ^{\prime }}\left( \mathbf{r}\right) =\frac{e^{2}}{\kappa
\left\vert \mathbf{r}-\mathbf{r}^{\prime }+\left( 1-\delta _{\xi ,\xi
^{\prime }}\right) d\widehat{\mathbf{z}}\right\vert }
\end{equation}%
has the Fourier transform%
\begin{equation}
V_{\xi ,\xi ^{\prime }}\left( \mathbf{r}\right) =\frac{1}{S}\sum_{\mathbf{q}}%
\frac{2\pi e^{2}}{\kappa q}e^{i\mathbf{q}\cdot \left( \mathbf{r}-\mathbf{r}%
^{\prime }\right) }e^{-qd\left( 1-\delta _{\xi ,\xi ^{\prime }}\right) },
\end{equation}%
where $\mathbf{q}$ is a two-dimensional vector in the plane of the bilayer.
The terms that do not conserve the valley index in Eq. (\ref{vcoulomb}) are
very small and usually neglected\cite{Goerbig1}.

\subsection{Hartree-Fock Hamiltonian}

In order to describe the different phases of the uniform C2DEG, we define
the operators%
\begin{equation}
\rho _{n,n^{\prime }}^{a,a^{\prime }}=\frac{1}{N_{\varphi }}%
\sum_{X}c_{a,n,X}^{\dagger }c_{a^{\prime },n^{\prime },X},
\end{equation}%
where $c_{a,n,X}^{\dagger }\left( c_{a,n,X}\right) $ creates(destroys) an
electron in a state $\left( a,n,X\right) $. The index $a$ combines the spin
and valley indices and we use $a_{\xi }$ and $a_{\sigma }$ to refer to the
specific spin or valley index of $a=\left( a_{\xi },a_{\sigma }\right) $.
The set of average values $\left\{ \left\langle \rho _{n,n^{\prime
}}^{a,a^{\prime }}\right\rangle \right\} $ gives a complete description of
an uniform ground state. They are the order parameters of that state. The
diagonal elements $\left\{ \left\langle \rho _{n,n}^{a,a}\right\rangle
\right\} $ are the filling factors of levels $\left( a,n\right) ,$ while the
off-diagonal elements are the "coherences". For nonuniform states, it is
necessary to define the order parameters $\left\{ \left\langle \rho
_{n,n^{\prime }}^{a,a^{\prime }}\left( \mathbf{G}\right) \right\rangle
\right\} $ where $\mathbf{G}$ is a reciprocal lattice vector and $%
\left\langle \rho _{n,n^{\prime }}^{a,a^{\prime }}\left( \mathbf{G}\right)
\right\rangle $ the Fourier transform of $\left\langle \rho _{n,n^{\prime
}}^{a,a^{\prime }}\left( \mathbf{r}\right) \right\rangle $. We refer the
reader to Refs. \onlinecite{CoteOrbital,CoteOrbital2} where the formalism
for this case is discussed in more details. The Hartree-Fock Hamiltonian can
be written in terms of these operators by (we adopt the convention that
repeated indices are summed over)

\begin{gather}
H_{HF}=N_{\varphi }E_{a,n}\rho _{n,n}^{a,a}  \label{HHF2} \\
-N_{\varphi }X_{n_{1},n_{4},n_{3},n_{2}}^{\left( a_{\xi },b_{\xi }\right)
}\left( 0\right) \left\langle \rho _{n_{1},n_{2}}^{a,b}\right\rangle \rho
_{n_{3},n_{4}}^{b,a},  \notag
\end{gather}%
where%
\begin{equation}
E_{a,n}=E_{a,n}^{0}+\alpha \left( \widetilde{\nu }_{a_{\xi }}-\frac{%
\widetilde{\nu }}{2}\right) \frac{d}{\ell },  \label{capa}
\end{equation}%
with $E_{a,n}^{0}$ given by Eqs. (\ref{s10}-\ref{s11}). The Fock interaction
is defined by%
\begin{eqnarray}
X_{n_{1},n_{2},n_{3},n_{4}}^{\left( a_{\xi },b_{\xi }\right) }\left( \mathbf{%
q}\right) &=&\alpha \int \frac{d\mathbf{p}\ell ^{2}}{2\pi }\frac{1}{p\ell }%
K_{n_{1},n_{2}}\left( \mathbf{p}\right) \\
&&\times K_{n_{3},n_{4}}\left( -\mathbf{p}\right) e^{i\mathbf{q}\times 
\mathbf{p}\ell ^{2}}e^{-pd\left( 1-\delta _{a_{\xi },b_{\xi }}\right) } 
\notag
\end{eqnarray}%
with the form factors%
\begin{eqnarray}
K_{0,0}\left( \mathbf{q}\right) &=&e^{-\frac{q^{2}\ell ^{2}}{4}},
\label{form1} \\
K_{1,1}\left( \mathbf{q}\right) &=&e^{-\frac{q^{2}\ell ^{2}}{4}}\left( 1-%
\frac{q^{2}\ell ^{2}}{2}\right) ,  \label{form2} \\
K_{1,0}\left( \mathbf{q}\right) &=&e^{-\frac{q^{2}\ell ^{2}}{4}}\left( \frac{%
\left( q_{y}+iq_{x}\right) \ell }{\sqrt{2}}\right) ,  \label{form3} \\
K_{0,1}\left( \mathbf{q}\right) &=&e^{-\frac{q^{2}\ell ^{2}}{4}}\left( \frac{%
\left( -q_{y}+iq_{x}\right) \ell }{\sqrt{2}}\right) .  \label{form4}
\end{eqnarray}%
These form factors capture the character of the two different orbital
states. In Eq. (\ref{capa}), $\widetilde{\nu }=\nu +4$ is the number of
filled levels in $N=0.$ We reserve the symbol $\nu \in \left[ -3,3\right] $
for the filling factor of the C2DEG. In deriving Eq. (\ref{HHF2}), we have
taken into account a neutralizing positive background so that the only
contribution from the Hartree and background terms is the capacitive energy
given by the term in parenthesis in Eq. (\ref{capa}). In this term, $%
\widetilde{\nu }_{a_{\xi }}=\sum_{n,\sigma }\left\langle \rho _{n,n}^{a_{\xi
},\sigma ;a_{\xi },\sigma }\left( 0\right) \right\rangle $ is the total
filling factor in valley $a_{\xi }.$ Detailed expressions for the Hartree
(see next section) and Fock interactions $H$ and $X$ are given in Appendix A
of Ref. \onlinecite{CoteOrbital}.

The Hartree-Fock energy per electron is given by%
\begin{gather}
\frac{E_{HF}}{N_{e}}=\frac{1}{\widetilde{\nu }}E_{a,n}^{0}\left\langle \rho
_{n,n}^{a,a}\right\rangle  \label{HFenergy} \\
+\frac{1}{4\widetilde{\nu }}\frac{d}{\ell }\alpha \left( \widetilde{\nu }%
_{K_{+}}-\widetilde{\nu }_{K_{-}}\right) ^{2}  \notag \\
-\frac{1}{2\widetilde{\nu }}X_{n_{1},n_{4},n_{3},n_{2}}^{\left( a_{\xi
},b_{\xi }\right) }\left( 0\right) \left\langle \rho
_{n_{1},n_{2}}^{a,b}\right\rangle \left\langle \rho
_{n_{3},n_{4}}^{b,a}\right\rangle ,  \notag
\end{gather}%
where $N_{e}$ is the number of electrons in the 2DEG and $\nu _{K_{\pm }}$
are the filling factors of the two valleys.

At $\mathbf{q}=0,$ the only nonzero matrix elements of the Fock interactions
are 
\begin{eqnarray}
X_{0,0,0,0}^{\xi ,\xi }\left( 0\right) &=&\Delta _{C},X_{1,1,1,1}^{\xi ,\xi
}\left( 0\right) =\frac{3}{4}\Delta _{C},  \label{v1} \\
X_{0,0,1,1}^{\xi ,\xi }\left( 0\right) &=&X_{1,1,0,0}^{\xi ,\xi }\left(
0\right) =\frac{1}{2}\Delta _{C},  \label{v2} \\
X_{1,0,0,1}^{\xi ,\xi }\left( 0\right) &=&X_{0,1,1,0}^{\xi ,\xi }\left(
0\right) =\frac{1}{2}\Delta _{C},  \label{v3}
\end{eqnarray}%
and the corresponding interlayer terms which must be computed numerically.
We have defined%
\begin{equation}
\Delta _{C}=\sqrt{\frac{\pi }{2}}\alpha
\end{equation}%
with $\alpha =e^{2}/\kappa \ell .$

\subsection{Calculation of the order parameters}

We define the single-particle Matsubara Green's function%
\begin{equation}
G_{n_{1,}n_{2}}^{a,b}\left( X,\tau \right) =-\left\langle T_{\tau
}c_{a,n_{1},X}\left( \tau \right) c_{b,n_{2},X}^{\dagger }\left( 0\right)
\right\rangle ,  \label{GFX}
\end{equation}%
where $T_{\tau }$ is the imaginary time ordering operator, such that the
order parameters are given by%
\begin{equation}
\left\langle \rho _{n_{1},n_{2}}^{a,b}\right\rangle =\frac{1}{N_{\varphi }}%
\sum_{X}G_{n_{2},n_{1}}^{b,a}\left( X,\tau =0^{-}\right) .
\end{equation}

The equation of motion for the Green's function in the Hartree-Fock
approximation is 
\begin{gather}
\left( i\hslash \omega _{n}+\mu -E_{a,n_{1}}\right)
G_{n_{1},n_{2}}^{a,b}\left( i\omega _{n}\right)  \label{EHF} \\
+U_{n_{1},n_{3}}^{a,c}G_{n_{3},n_{2}}^{c,b}\left( i\omega _{n}\right)
=\hslash \delta _{n_{1},n_{2}}\delta _{a,b},  \notag
\end{gather}%
where $\mu $ is the chemical potential, $\omega _{n}$ a fermionic Matsubara
frequency and%
\begin{equation}
U_{n_{1},n_{3}}^{a,c}=X_{n_{4},n_{3},n_{1},n_{2}}^{\left( a_{\xi },c_{\xi
}\right) }\left( 0\right) \left\langle \rho _{n_{4},n_{2}}^{c,a}\right\rangle
\end{equation}%
are the self-consistent Fock potentials.

The self-consistent Eq. (\ref{EHF}) can be put in a $8\times 8$ matrix form
by defining superindices and then solved numerically in an iterative way in
order to get the order parameters.

The Hartree-Fock equation of motion for the Green's function leads to the
sum rule (at $T=0$ K)%
\begin{equation}
\sum_{b,m}\left\vert \left\langle \rho _{n,m}^{a,b}\right\rangle \right\vert
^{2}=\left\langle \rho _{n,n}^{a,a}\right\rangle =\nu _{n}^{a},  \label{SR}
\end{equation}%
where $\nu _{n}^{a}$ is the filling factor of the $\left( a,n\right) $
level. By definition%
\begin{equation}
\left\langle \rho _{n,m}^{a,b}\right\rangle =\left\langle \rho
_{m,n}^{b,a}\right\rangle ^{\ast }.
\end{equation}

\subsection{Collective modes in the generalized random-phase approximation}

To study the collective excitations, we compute the two-particle Green's
functions

\begin{align}
& \chi _{n_{1},n_{2},n_{3},n_{4}}^{a,b,c,d}\left( \mathbf{q},\tau \right)
\label{twopart} \\
& =-N_{\varphi }\left\langle T_{\tau }\rho _{n_{1},n_{2}}^{a,b}\left( 
\mathbf{q,}\tau \right) \rho _{n_{3},n_{4}}^{c,d}\left( -\mathbf{q},0\right)
\right\rangle  \notag \\
& +N_{\varphi }\left\langle \rho _{n_{1},n_{2}}^{a,b}\left( \mathbf{q}%
\right) \right\rangle \left\langle \rho _{n_{3},n_{4}}^{c,d}\left( -\mathbf{q%
}\right) \right\rangle  \notag
\end{align}%
in the generalized random-phase approximation (GRPA). In this approximation, 
$\chi _{n_{1},n_{2},n_{3},n_{4}}^{a,b,c,d}\left( \mathbf{q},\tau \right) $
is the solution of the equation

\begin{eqnarray}
&&\chi _{n_{1},n_{2},n_{3},n_{4}}^{a,b,c,d}\left( \mathbf{q},i\Omega
_{n}\right)  \label{grpa1} \\
&=&\chi _{n_{1},n_{2},n_{3},n_{4}}^{\left( 0\right) a,b,c,d}\left( \mathbf{q}%
,i\Omega _{n}\right)  \notag \\
&&+\frac{1}{\hslash }\chi _{n_{1},n_{2},n_{5},n_{6}}^{\left( 0\right)
a,b,e,e}\left( \mathbf{q},i\Omega _{n}\right)  \notag \\
&&\times H_{n_{5},n_{6},n_{7},n_{8}}^{\left( e_{\xi },g_{\xi }\right)
}\left( \mathbf{q}\right) \chi _{n_{7},n_{8},n_{3},n_{4}}^{g,g,c,d}\left( 
\mathbf{q},i\Omega _{n}\right)  \notag \\
&&-\frac{1}{\hslash }\chi _{n_{1},n_{2},n_{5},n_{6}}^{\left( 0\right)
a,b,e,f}\left( \mathbf{q},i\Omega _{n}\right)  \notag \\
&&\times X_{n_{5},n_{8},n_{7},n_{6}}^{\left( e_{\xi },f_{\xi }\right)
}\left( \mathbf{q}\right) \chi _{n_{7},n_{8},n_{3},n_{4}}^{f,e,c,d}\left( 
\mathbf{q},i\Omega _{n}\right) ,  \notag
\end{eqnarray}%
where $\Omega _{n}$ is a bosonic Matsubura frequency and the Hartree
interaction%
\begin{eqnarray}
H_{n_{1},n_{2},n_{3},n_{4}}^{\left( a_{\xi },b_{\xi }\right) }\left( \mathbf{%
q}\right) &=&\frac{1}{q\ell }K_{n_{1},n_{2}}\left( \mathbf{q}\right)
K_{n_{3},n_{4}}\left( -\mathbf{q}\right) \\
&&\times e^{-qd\left( 1-\delta _{a_{\xi },b_{\xi }}\right) }.  \notag
\end{eqnarray}%
The two-particle Green's functions $\chi _{n_{1},n_{2},n_{3},n_{4}}^{\left(
0\right) a,b,c,d}\left( \mathbf{q},i\Omega _{n}\right) $ satisfy the set of
equations

\begin{eqnarray}
&&\left[ i\hslash \Omega _{n}-\left( E_{b,n_{2}}-E_{a,n_{1}}\right) \right]
\chi _{n_{1},n_{2},n_{3},n_{4}}^{\left( 0\right) a,b,c,d}\left( \mathbf{q}%
,\Omega _{n}\right)  \label{grpa2} \\
&=&\hslash \left\langle \rho _{n_{1},n_{4}}^{a,d}\right\rangle \delta
_{b,c}\delta _{n_{2},n_{3}}-\hslash \left\langle \rho
_{n_{3},n_{2}}^{c,b}\right\rangle \delta _{a,d}\delta _{n_{1},n_{4}}  \notag
\\
&&+U_{m,n_{1}}^{a,e}\chi _{m,n_{2},n_{3},n_{4}}^{\left( 0\right)
e,b,c,d}\left( \mathbf{q},\Omega _{n}\right)  \notag \\
&&-U_{n_{2},m}^{e,b}\chi _{n_{1},m,n_{3},n_{4}}^{\left( 0\right)
a,e,c,d}\left( \mathbf{q},\Omega _{n}\right) .  \notag
\end{eqnarray}%
Eq. (\ref{grpa1}) can be represented by a set of bubbles (Hartree terms) and
ladder (Fock terms) diagrams. The function $\chi
_{n_{1},n_{2},n_{3},n_{4}}^{\left( 0\right) a,b,c,d}$ is the Hartree-Fock
approximations for the two-particle Green's functions. It includes the
Hartree-Fock self-energy corrections but not the vertex corrections. Note
that two-particle Green's functions depend only on the order parameters $%
\left\langle \rho _{n,m}^{a,b}\right\rangle $ computed in the HFA. Eqs. (\ref%
{grpa1},\ref{grpa2}) can be solved numerically by defining superindices and
then writing them in a $64\times 64$ matrix form. The collective excitations
are then given by the poles of the retarded Green's functions $\chi
_{n_{1},n_{2},n_{3},n_{4}}^{\left( R\right) a,b,c,d}\left( \mathbf{q},\omega
\right) $ which are obtained by the analytic continuation $i\Omega
_{n}\rightarrow \omega +i\delta $ of the corresponding two-particle Green's
functions. To derive the dispersion relations, we follow these poles as the
wave vector $\mathbf{q}$ is varied.

\subsection{Pseudospin description}

We showed above that the coherent states of the C2DEG can be described by
the set of order parameters $\left\{ \left\langle \rho _{n,n^{\prime
}}^{a,a^{\prime }}\right\rangle \right\} .$ These states are also quantum
Hall ferromagnets (QHF's) and can also be described by using a pseudospin
language where the two valley states $\xi =+$ ($\xi =-$) are associated with
valley-pseudospin up (down) and the two orbital states $n=0$ ($n=1$) with
orbital-pseudospin up (down).

In this language, the total spin, valley pseudospin, and orbital pseudospin
components of the electron gas are given by%
\begin{eqnarray}
S_{i} &=&\frac{1}{2N_{\varphi }}\hslash \sum_{\xi ,n,X}\sum_{\alpha ,\beta
}\left\langle c_{\xi ,\alpha ,n,X}^{\dag }\sigma _{\alpha ,\beta }^{\left(
i\right) }c_{\xi ,\beta ,n,X}\right\rangle , \\
L_{i} &=&\frac{1}{2N_{\varphi }}\sum_{\alpha ,n,X}\sum_{\xi ,\xi ^{\prime
}}\left\langle c_{\xi ,\alpha ,n,X}^{\dag }\sigma _{\xi ,\xi ^{\prime
}}^{\left( i\right) }c_{\xi ^{\prime },\alpha ,n,X}\right\rangle , \\
O_{i} &=&\frac{1}{2N_{\varphi }}\sum_{\xi ,\alpha ,X}\sum_{n,n^{\prime
}}\left\langle c_{\xi ,\alpha ,n,X}^{\dag }\sigma _{n,n^{\prime }}^{\left(
i\right) }c_{\xi ,\alpha ,n^{\prime },X}\right\rangle ,
\end{eqnarray}%
where $\sigma ^{\left( i\right) \prime }s$ are the Pauli matrices and the
total filling factor is%
\begin{equation}
\widetilde{\nu }=\frac{1}{N_{\varphi }}\sum_{\xi ,\alpha ,n,X}\left\langle
c_{\xi ,\alpha ,n,X}^{\dag }c_{\xi ,\alpha ,n,X}\right\rangle .
\end{equation}%
Note that these $10$ fields do not provide a complete description of a
state. One must also consider the $54$ other combinations of indices (the $%
64 $ order parameters are not all independent, however).We will use both $%
S_{i},L_{i},O_{i}$ and the order parameters $\left\langle \rho
_{n,m}^{a,b}\right\rangle $ to characterize the ground states of the C2DEG.

\subsection{Induced dipoles}

The coupling of the C2DEG with a uniform external electric field \textit{in
the plane of the layers} is given by 
\begin{equation}
H_{E}=-e\int d\mathbf{r}n\left( \mathbf{r}\right) \phi \left( \mathbf{r}%
\right)  \label{couplage}
\end{equation}%
where $\mathbf{E}_{ext}=-\nabla \phi \left( \mathbf{r}\right) .$ The total
density is given by%
\begin{equation}
n\left( \mathbf{r}\right) =\sum_{\sigma ,\xi ,n,m}\Psi _{\xi ,\sigma
,n}^{\dag }\left( \mathbf{r}\right) \Psi _{\xi ,\sigma ,m}\left( \mathbf{r}%
\right) .
\end{equation}%
Fourier transforming Eq. (\ref{couplage}) and using the form factors defined
in Eqs. (\ref{form1}-\ref{form4}), we can show that in an homogeneous state%
\cite{Shizuyadipole,CoteOrbital}, $H_{E}$ gives the dipolar coupling

\begin{equation}
H_{E}=-\mathbf{d}\cdot \mathbf{E}_{ext},
\end{equation}%
with the total dipole vector defined by $\mathbf{d}=\sum_{a}\mathbf{d}_{a}$
where 
\begin{equation}
\mathbf{d}_{a}=-\sqrt{2}\ell eN_{\varphi }\left( \rho _{x}^{a,a}\left(
0\right) \widehat{\mathbf{x}}-\rho _{y}^{a,a}\left( 0\right) \widehat{%
\mathbf{y}}\right)  \label{dipole}
\end{equation}%
is the dipole moment in valley $a_{\xi }$ with spin $a_{\sigma }.$ We have
defined here%
\begin{eqnarray}
\rho _{x}^{a} &=&\frac{1}{2}\left( \rho _{0,1}^{a,a}+\rho
_{1,0}^{a,a}\right) ,  \label{cun} \\
\rho _{y}^{a} &=&\frac{1}{2i}\left( \rho _{0,1}^{a,a}-\rho
_{1,0}^{a,a}\right) .  \label{cdeux}
\end{eqnarray}%
It is possible to control the orientation of the orbital pseudospins in the $%
x-y$ plane with an external electric field.

\subsection{Electromagnetic absorption}

The total current operator in second quantization is given by $\mathbf{J}%
=\sum_{a}\mathbf{J}_{a}$with 
\begin{equation}
\mathbf{J}_{a}=\sum_{n,m}\int d\mathbf{r}\Psi _{a,n}^{\dag }\left( \mathbf{r}%
\right) \mathbf{j}_{a}\left( \mathbf{r}\right) \Psi _{a,m}\left( \mathbf{r}%
\right) ,
\end{equation}%
The current operator $\mathbf{j}_{a}\left( \mathbf{r}\right) $ is derived
from the Hamiltonian in Eq. (\ref{hamilp}) by making the Peierls
substitution $\mathbf{p}\rightarrow \mathbf{P}=\mathbf{p}+e\mathbf{A}%
_{ext}/c $ and then taking the derivative with respect to the external
vector potential $\mathbf{A}_{ext}$ 
\begin{equation}
\mathbf{j}_{a,i}=-c\left. \frac{\partial H_{a}^{0}}{\partial A_{i}^{e}}%
\right\vert _{A_{i}^{e}\rightarrow 0}
\end{equation}%
where $i=x,y.$ This gives%
\begin{equation}
\mathbf{J}=\frac{1}{\hslash }\sum_{a}\Delta _{a}^{0}\left( \widehat{\mathbf{z%
}}\times \mathbf{d}_{a}\right) =\frac{d\mathbf{d}}{dt},
\end{equation}%
with%
\begin{equation}
\Delta _{a}^{0}=E_{a,1}^{0}-E_{a,0}^{0}=\zeta _{1}+a_{\xi }\beta \Delta _{B}
\label{xsip}
\end{equation}%
i.e. the bare gap in valley $\xi $ and%
\begin{equation}
\frac{d\mathbf{d}}{dt}=-\frac{i}{\hslash }\left[ H_{HF}^{0},\mathbf{d}\right]
,
\end{equation}%
where $H_{HF}^{0}=N_{\varphi }E_{a,n}^{0}\rho _{n,n}^{a,a}$ is the
noninteracting Hamiltonian.

To compute the electromagnetic absorption per unit area, we define the
two-particle current-current Green's function%
\begin{equation}
\chi _{J_{\alpha },J_{\beta }}\left( \tau \right) =-\frac{1}{S}\left\langle
T_{\tau }J_{\alpha }\left( \tau \right) J_{\beta }\left( 0\right)
\right\rangle ,
\end{equation}%
which gives

\begin{eqnarray}
\chi _{J_{\alpha },J_{\beta }}\left( i\Omega _{n}\right) &=&\left( \frac{%
e\ell }{\hslash }\right) ^{2}\frac{1}{\pi \ell ^{2}}\sum_{a,b}\Delta
_{a}^{0}\Delta _{b}^{0}  \label{absoabso} \\
&&\times \chi _{\rho _{\overline{\alpha }},\rho _{\overline{\beta }%
}}^{a,a,b,b}\left( \mathbf{q}=0,i\Omega _{n}\right) ,  \notag
\end{eqnarray}%
where $\alpha ,\beta =x,y$ and $\overline{x}=y$,$\overline{y}=x.$ Using Eqs.
(\ref{cun},\ref{cdeux}), we have for example%
\begin{equation}
\chi _{\rho _{x},\rho _{x}}^{a,b,c,d}=\frac{1}{4}\left( \chi
_{1,0,1,0}^{a,b,c,d}+\chi _{0,1,1,0}^{a,b,c,d}+\chi
_{1,0,0,1}^{a,b,c,d}+\chi _{0,1,0,1}^{a,b,c,d}\right)
\end{equation}%
and similarly for the other components. The absorption can only involve
these four combinations of orbital indices whatever the polarization of the
electric field of the electromagnetic wave. The retarded current-current
response function $\chi _{J_{\alpha },J_{\beta }}\left( \omega \right) $ is
obtained from the analytic continuation $i\Omega _{n}\rightarrow \omega
+i\delta $ and the electromagnetic absorption for an electromagnetic wave of
amplitude $E_{0}$ linearly polarized in the direction $\alpha $ is given by%
\begin{equation}
P_{\alpha }\left( \omega \right) =-\frac{1}{\hslash }\Im\left[ \frac{%
\chi _{J_{\alpha },J_{\alpha }}\left( \omega \right) }{\omega }\right]
E_{0}^{2}.  \label{abso}
\end{equation}%
This formula is valid at finite frequency only since we have neglected the
diamagnetic contribution to the current.

\subsection{Absorption in the incoherent phases}

If there is no coherence in a phase, then 
\begin{equation}
\left\langle \rho _{n,m}^{a,b}\right\rangle =\left\langle \rho
_{n,n}^{a,a}\right\rangle \delta _{n,m}\delta _{a,b}.
\end{equation}%
In this case, we can solve analytically for the absorption because this
restriction leads, from Eq. (\ref{grpa2}) to%
\begin{eqnarray}
&&\chi _{n_{1},n_{2},n_{3},n_{4}}^{\left( 0\right) a,b,c,d}\left( \mathbf{q}%
,\Omega _{n}\right) \\
&=&\chi _{n_{1},n_{2},n_{2},n_{1}}^{\left( 0\right) a,b,b,a}\left( \mathbf{q}%
,\Omega _{n}\right) \delta _{a,d}\delta _{b,c}\delta _{n_{1},n_{4}}\delta
_{n_{2},n_{3}}.  \notag
\end{eqnarray}%
Now, at $\mathbf{q}=0$ the only nonzero Fock interactions in the GRPA\
equations are given in Eqs. (\ref{v1}-\ref{v3}) while the only Hartree
interactions that need to be considered are those of the form 
\begin{equation}
H_{0,0,0,0}^{\xi ,-\xi }\left( 0\right) ,H_{1,1,1,1}^{\xi ,-\xi }\left(
0\right) ,H_{0,0,1,1}^{\xi ,-\xi }\left( 0\right) ,H_{1,1,0,0}^{\xi ,-\xi
}\left( 0\right) .
\end{equation}%
These interlayer Hartree interaction involves combinations of the form $%
e^{-qd}/q$ that give a finite contribution at $\mathbf{q}=0$ (and also a
diverging contribution that is cancelled by the other terms). It follows
that Eq. (\ref{grpa1}) gives for the GRPA response functions $\chi
_{1,0,1,0}^{a,a,c,c}\left( \omega \right) =\chi _{0,1,0,1}^{a,a,c,c}\left(
\omega \right) =0$ and 
\begin{equation}
\chi _{0,1,1,0}^{a,a,c,c}\left( \omega \right) =\frac{\chi
_{0,1,1,0}^{\left( 0\right) a,a,a,a}\left( \omega \right) \delta _{a,c}}{%
\left[ 1+\frac{1}{\hslash }X_{1,1,0,0}^{\left( a_{\xi },a_{\xi }\right)
}\left( 0\right) \chi _{0,1,1,0}^{\left( 0\right) a,a,a,a}\left( \omega
\right) \right] },
\end{equation}%
i.e. valley and spin must be conserved in an optically active electronic
transition. Since 
\begin{equation}
\chi _{0,1,1,0}^{\left( 0\right) a,a,a,a}\left( \omega \right) =\frac{%
\left\langle \rho _{0,0}^{a,a}\right\rangle -\left\langle \rho
_{1,1}^{a,a}\right\rangle }{\omega +i\delta -\left( \Delta
_{a}^{0}+U_{0,0}^{a,a}-U_{1,1}^{a,a}\right) /\hslash }
\end{equation}%
(and a similar expression with $0\rightleftharpoons 1$ for $\chi
_{1,0,0,1}^{\left( 0\right) a,a,a,a}\left( \omega \right) $), we have easily%
\begin{equation}
\chi _{0,1,1,0}^{a,a,a,a}\left( \omega \right) =\frac{\left\langle \rho
_{0,0}^{a,a}\right\rangle -\left\langle \rho _{1,1}^{a,a}\right\rangle }{%
\omega +i\delta -\left[ \Delta _{a}^{0}+\frac{\Delta _{C}}{4}\left\langle
\rho _{1,1}^{a,a}\right\rangle \right] /\hslash }
\end{equation}%
and%
\begin{equation}
\chi _{1,0,0,1}^{a,a,a,a}\left( \omega \right) =\frac{\left\langle \rho
_{1,1}^{a,a}\right\rangle -\left\langle \rho _{0,0}^{a,a}\right\rangle }{%
\omega +i\delta +\left[ \Delta _{a}^{0}+\frac{\Delta _{C}}{4}\left\langle
\rho _{1,1}^{a,a}\right\rangle \right] /\hslash }.
\end{equation}%
The functions $\chi _{0,1,1,0}^{a,a,a,a}\left( \omega \right) $ and $\chi
_{1,0,0,1}^{a,a,a,a}\left( \omega \right) $ are the response to the two
circular polarizations of light. The absorption in an incoherent phase is
finally given by%
\begin{eqnarray}
P_{\alpha }\left( \omega \right) &=&\frac{E_{0}^{2}}{4\ell ^{2}\omega }%
\left( \frac{e\ell }{\hslash }\right) ^{2}\sum_{a}\left( \left\langle \rho
_{0,0}^{a,a}\right\rangle -\left\langle \rho _{1,1}^{a,a}\right\rangle
\right) \\
&&\times \left( \Delta _{a}^{0}\right) ^{2}\delta \left( \hslash \omega
-\Delta _{a}^{0}-\frac{\Delta _{C}}{4}\left\langle \rho
_{1,1}^{a,a}\right\rangle \right)  \notag
\end{eqnarray}%
with $\alpha =x,y$ and $\Delta _{a}^{0}$ given by Eq. (\ref{xsip}). In the
numerical calculation, we introduce a small Landau level width in order to
get a finite value for the optical absorption.

We can follow the same type of reasoning to show that, in a phase with no
orbital coherence but with possibly layer and/or spin coherence, the
functions $\chi _{1,0,1,0}^{a,b,c,d}\left( \omega \right) =\chi
_{0,1,0,1}^{a,b,c,d}\left( \omega \right) =0$ and the absorption depends
again only on $\chi _{0,1,1,0}^{a,a,b,b}\left( \omega \right) $ and $\chi
_{1,0,0,1}^{a,a,b,b}\left( \omega \right) .$ In this special case, the
equation of motion for $\chi _{0,1,1,0}^{a,b,c,d}\left( \omega \right) $ is%
\begin{eqnarray}
\chi _{0,1,1,0}^{a,a,b,b}\left( \omega \right) &=&\chi _{0,1,1,0}^{\left(
0\right) a,a,b,b}\left( \omega \right)  \label{chizero} \\
&&-\frac{1}{\hslash }\sum_{e,f}\chi _{0,1,1,0}^{\left( 0\right)
a,a,e,f}\left( \omega \right) X_{1,1,0,0}^{e_{\xi },f_{\xi }}\chi
_{0,1,1,0}^{f,e,b,b}\left( \omega \right)  \notag
\end{eqnarray}%
and a similar expression with $0\rightleftharpoons 1$ for $\chi
_{1,0,0,1}^{a,b,c,d}\left( \omega \right) .$

\section{PHASE DIAGRAM OF THE C2DEG}

At zero bias, the QHF states follow a set of Hund's rules: the spin
polarization is maximized first, then the layer polarization is maximized to
the greatest extent possible and finally the orbital polarization is
maximized to the extent allowed by the first two rules\cite{BarlasPRL1}. In
this section, we study the phase transitions that occur when a \textit{finite%
} bias (or transverse electric field) is turned on.

\begin{figure*}[tbph]
\includegraphics[scale=0.7]{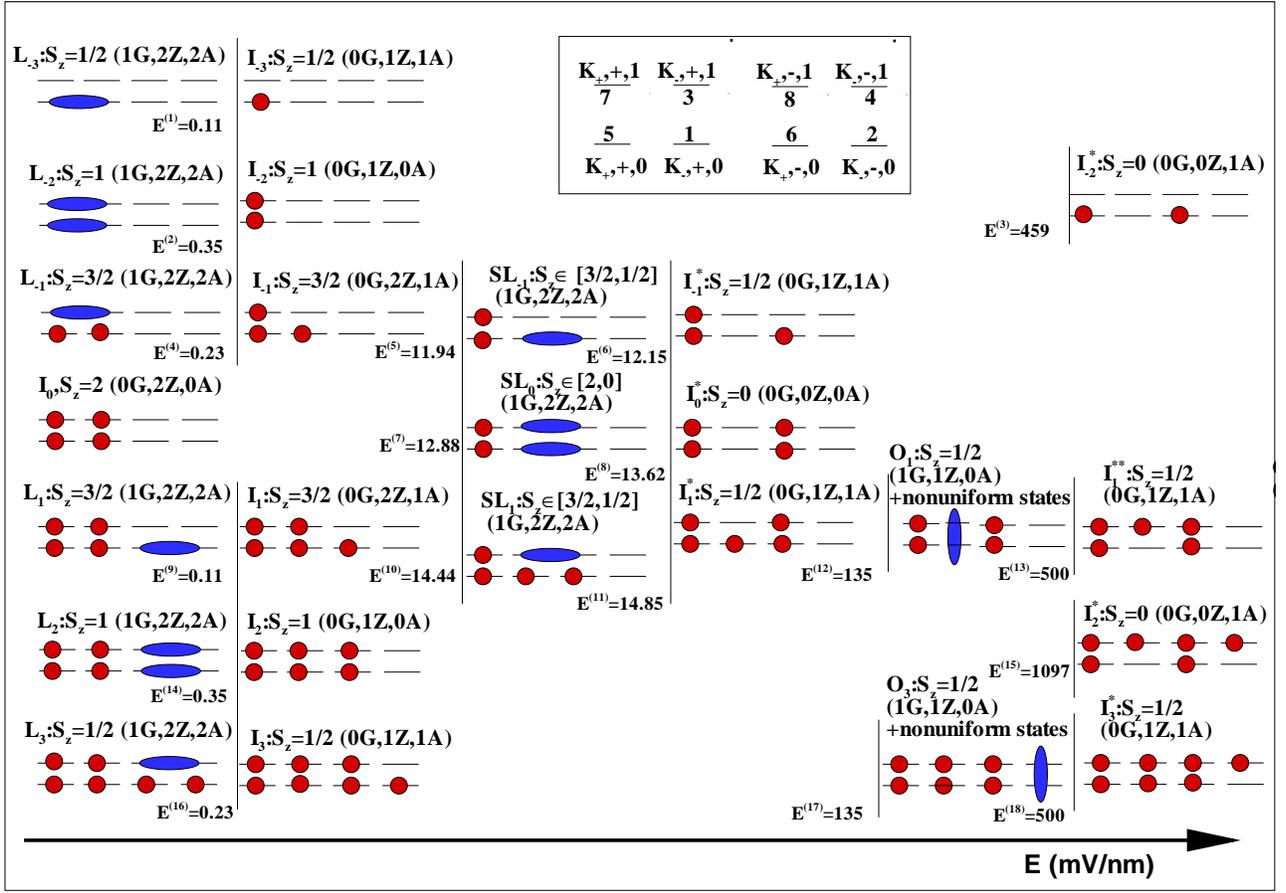}
\caption{(Color online) Phase diagram of the C2DEG in Landau level $N=0$ at $%
B=10$ T and for $\protect\kappa =5$ as a function of the transverse electric
field beween the layers for integer filling factors $\protect\nu \in \left[
-3,3\right] $ (from top to bottom). A filled (red) circle represents a
filled state while a filled (blue) ellipse indicates a coherent
superposition of two levels. The numbering of the levels is indicated in the
inset at the top right of the figure. The $E^{(i)\prime }s$ indicate the
critical perpendicular electric field (in mV/nm) required for the transition
between two phases. Also indicated for each phase are the spin polarization $%
S_{z},$ the number of Goldstone mode (G), of modes gapped at the Zeeman
energy (Z) and the number of peaks in the optical absorption spectrum (A).
The mention "+ nonuniform states" for the $O_{1}$ and $O_{3}$ phases signals
that this portion of the phase diagram is further subdivided into uniform
and nonuniform states as indicated in Eq. (\protect\ref{nonphase}).}
\label{figdiagrammeglobal}
\end{figure*}

\subsection{Types of phases}

Fig. \ref{figdiagrammeglobal} shows our numerical result for the phase
diagram of the C2DEG as a function of an applied transverse electric field $%
E=\Delta _{B}/ed$ for $B=10$ T and $\kappa =5$ and for all integer filling
factors $\nu \in \left[ -3,3\right] .$ We indicate the eight non-interacting
levels by horizontal lines and number them according to the scheme indicated
in the top inset. Note that the lines are only offset vertically for
clarity. Their position does not reflect the true ordering of the energy
levels which changes with bias. We name the phases $I_{\nu },L_{\nu },O_{\nu
}$ and $SL_{\nu }$ according to the type of coherence that is present:
incoherent, layer-coherent, orbital-coherent or spin-layer coherent
respectively. When there is more than one incoherent phase at a given
filling factor, we use the notation $I_{\nu }^{\ast }$ for the second phase, 
$I_{\nu }^{\ast \ast }$ for the third phase and so on. The critical electric
field for the transition between two phases is indicated by $E^{(1)}$ to $%
E^{(18)}$ and is in units of mV/nm. A circle on an energy level represents a
fully filled level while an ellipse that connects two levels indicates a
coherent superposition of these two states. We list in Fig. \ref%
{figdiagrammeglobal} some properties of each phase: spin polarization $%
S_{z}, $ number of Goldstone modes (G), number of collective modes gapped at
the Zeeman energy (Z)\ and number of peaks in the optical absorption
spectrum (A). The transition between a coherent and an incoherent phase is
continuous while a transition between two incoherent phases is discontinuous.

\subsubsection{Incoherent phases $I_{\protect\nu }$}

The $I_{\nu }$ phases have no coherence of any kind and so $L_{\Vert
},S_{\Vert },O_{\Vert }=0$ (the parallel component is in the plane of the
bilayer). Each level is either full or empty and so $L_{z},S_{z}$ and $O_{z}$
vary from one phase to another. The state corresponding to a specific
diagram is easily read from Fig. 5. For $I_{2}$, we have for example%
\begin{equation}
\left\vert \Psi _{I_{2}}\right\rangle =\prod\limits_{X}c_{7,X}^{\dag
}c_{5,X}^{\dag }\left\vert 0\right\rangle ,
\end{equation}%
and the order parameters $\left\langle \rho _{5,5}\right\rangle
=\left\langle \rho _{7,7}\right\rangle =1.$ We include the phases $I_{\pm
2}^{\ast },I_{1}^{\ast \ast }$ and $I_{3}^{\ast }$ in the phase diagram only
to make it more complete. Indeed, the bias can't produce any more transition
after these states. But, the critical bias needed to reach these states is
well outside the limits of validity of our two-band model.

Using Eq. (\ref{HFenergy}), the Hartree-Fock energy of two adjacent
incoherent phases are readily compared to extract the critical biases. We
find, with $\Delta _{B}^{\left( i\right) }=edE^{\left( i\right) },$ 
\begin{eqnarray}
I_{-2} &\rightarrow &I_{-2}^{\ast }:\Delta _{B}^{\left( 3\right) }=\frac{%
\Delta _{Z}-\zeta _{1}+\frac{3}{8}\Delta _{C}}{\beta }, \\
I_{+2} &\rightarrow &I_{+2}^{\ast }:\Delta _{B}^{\left( 15\right) }=\frac{%
\Delta _{Z}+\zeta _{1}+\frac{5}{8}\Delta _{C}}{\beta }.
\end{eqnarray}%
If we ignore the $SL_{\nu }$ phases at $\nu =0,\pm 1$, we find that the
transition between the two incoherent phases occurs in the middle of the $%
SL_{\nu }$ phase (see Sec. IV.A.4 below) i.e. at 
\begin{eqnarray}
I_{-1} &\rightarrow &I_{-1}^{\ast }:\Delta _{-1}=\Delta _{z}+2\frac{d}{\ell }%
\alpha ,  \label{gapm1} \\
I_{0} &\rightarrow &I_{0}^{\ast }:\Delta _{0}=\frac{\Delta _{z}+2\frac{d}{%
\ell }\alpha }{1-\beta },  \label{gap0} \\
I_{1} &\rightarrow &I_{1}^{\ast }:\Delta _{1}=\frac{\Delta _{z}+2\frac{d}{%
\ell }\alpha }{1-2\beta }.  \label{gapp1}
\end{eqnarray}%
Note that these last three results are independent of Coulomb exchange
corrections and so of screening corrections. (The capacitive term comes from
the Hartree self-energy and is not screened.)

From Eq. (\ref{gap0}), we find a critical electric field $E_{c}=\Delta
_{0}/ed\approx 4.7B\left[ \text{T}\right] $ mV/nm when $\kappa =1$ for the
transition $I_{0}\rightarrow I_{0}^{\ast }$. This critical field depends
linearly on the magnetic field, in agreement with the experiments\cite{Weitz}%
. Experimentally, however, the slope is \cite{Weitz} $11$ mV nm$^{-1}$ T$%
^{-1}$ or\cite{Velasco} $12.7$ mV nm$^{-1}$ T$^{-1}$ or\cite{Kim} $12-18$ mV
nm$^{-1}$ T$^{-1}$ and thus larger than the HFA value. Moreover, experiments
measure an offset of $E_{c}\approx 20$ mV/nm at $B=0$ T\cite{Weitz}. This
offset can't be captured by our HFA which is only valid at sufficiently
large magnetic field where Landau level mixing can be neglected. Apart from
the extra $\beta $ corrections, Eqs. (\ref{gapm1}-\ref{gapp1}) are identical
to those given by Gorbar et al.\cite{Gorbarnutot}.

\subsubsection{Layer-coherent phases $L_{\protect\nu }$}

\begin{figure}[tbph]
\includegraphics[scale=0.8]{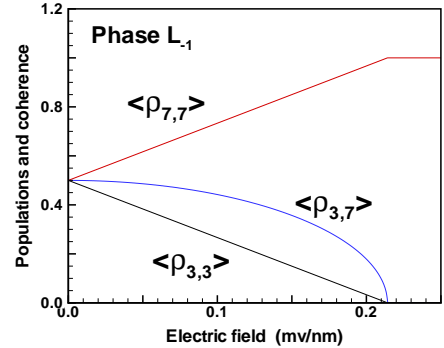}
\caption{(Color online) Variation of the populations and interlayer
coherence with the transverse electric field in phase $L_{-1}.$ }
\label{figpopLm1}
\end{figure}

The second type of phase, $L_{\nu }$, has layer coherence between two states
with the same spin and orbital indices and so $L_{\Vert }\neq 0$ but $%
S_{\Vert },O_{\Vert }=0.$ The tilt angle of the pseudospin vector $\mathbf{L}
$ varies with bias in this phase but $O_{z}$ and $S_{z}$ are constant. An
example is phase $L_{-3}$ which is described by 
\begin{equation}
\left\vert \Psi _{L_{-3}}\right\rangle =\prod\limits_{X}\left(
ac_{5,X}^{\dag }+bc_{1,X}^{\dag }\right) \left\vert 0\right\rangle ,
\end{equation}%
where the coefficients $a$ and $b$ depend on the bias and are related by $%
\left\vert a\right\vert ^{2}+\left\vert b\right\vert ^{2}=1.$ With
increasing bias, $a\rightarrow 1$ and $b\rightarrow 0$ continuously. The
level populations and the coherence in $L_{-3}$ are given by 
\begin{eqnarray}
\left\langle \rho _{5,5}\right\rangle &=&\left\vert a\right\vert
^{2},\left\langle \rho _{1,1}\right\rangle =\left\vert b\right\vert ^{2}, \\
\left\langle \rho _{1,5}\right\rangle &=&\left\langle \rho
_{5,1}\right\rangle ^{\ast }=ab^{\ast }.
\end{eqnarray}%
Fig. \ref{figpopLm1} shows how these variables depend on the transverse
electric field for the similar phase $L_{-1}.$ The populations of the
coherent levels vary linearly with the bias in all $L_{\nu }$ phases with
the exception of $L_{\pm 2}$ where the variation is not exactly linear. In $%
L_{-1},$ for example, 
\begin{equation}
\left\langle \rho _{5,5}\right\rangle =\frac{1}{2}\left( 1+\frac{\Delta _{B}%
}{\Delta _{B}^{(1)}}\right) ,
\end{equation}%
where the critical bias is, to order $\left( d/\ell \right) ^{2},$%
\begin{eqnarray}
L_{-3\left( 1\right) } &\rightarrow &I_{-3\left( 1\right) }:\Delta
_{B}^{(1)}=\Delta _{B}^{(9)}\approx \sqrt{\frac{\pi }{8}}\left( \frac{d}{%
\ell }\right) ^{2}\alpha , \\
L_{-1\left( 3\right) } &\rightarrow &I_{-1\left( 3\right) }:\Delta
^{(4)}=\Delta _{B}^{(16)} \\
&\approx &\frac{7}{4}\sqrt{\frac{\pi }{8}}\left( \frac{d}{\ell }\right)
^{2}\alpha .  \notag
\end{eqnarray}%
These critical biases all scale with the magnetic field as $B^{3/2}.$

In phases $L_{\pm 2},$ there is a layer coherence in orbitals $n=0$ and $%
n=1. $ Phase $L_{-2},$ for example, is described by the state%
\begin{eqnarray}
\left\vert \Psi _{L_{-2}}\right\rangle &=&\prod\limits_{X}\left( \alpha
c_{5,X}^{\dag }+\gamma c_{1,X}^{\dag }\right) \\
&&\times \left( \alpha ^{\prime }c_{7,X}^{\dag }+\gamma ^{\prime
}c_{3,X}^{\dag }\right) \left\vert 0\right\rangle .  \notag
\end{eqnarray}%
The critical bias $\Delta ^{(2)}=\Delta ^{(14)}$ for the transitions $L_{\pm
2}\rightarrow I_{\pm 2}$ has a complicated analytical expression that we do
not reproduce here but the numerical values of the critical electric field
for $B=10$ T and $\kappa =5$ is indicated in Fig. \ref{figdiagrammeglobal}.

Phases with layer coherence occur in a small range of bias and at very small
bias because the interlayer separation $d/\ell =0.013\sqrt{B}$ is very small
in bilayer graphene and so is the capacitive energy. In semiconductor
bilayers, $d/\ell $ can be of order unity and interlayer coherence can
survive to a much higher bias\cite{Ezawa}.

\subsubsection{Orbital-coherent phases $O_{\protect\nu }$}

For $\Delta _{B}>\Delta _{B}^{\left( 12\right) }=\Delta _{B}^{\left(
17\right) },$ the ordering of the energy levels $n=0,1$ is reversed (this
change is not shown in Fig. \ref{figdiagrammeglobal}). When this happens,
the kinetic energy is minimized by filling level $n=1$ before $n=0.$
However, this increases the Coulomb exchange energy because $%
X_{1,1,1,1}^{+,+}\left( 0\right) <X_{0,0,0,0}^{+,+}\left( 0\right) .$ The
C2DEG optimizes its energy by creating a coherent superposition of $n=0$ and 
$n=1$ with the same valley and spin indices. We use the notation $O_{\nu }$
for such a phase. An example is phase $O_{3}$ which is described by 
\begin{eqnarray}
\left\vert \Psi _{O_{3}}\right\rangle &=&\prod\limits_{X}\left(
ac_{2,X}^{\dag }+bc_{4,X}^{\dag }\right) c_{8,X}^{\dag }c_{6,X}^{\dag } \\
&&\times c_{7,X}^{\dag }c_{5,X}^{\dag }c_{3,X}^{\dag }c_{1,X}^{\dag
}\left\vert 0\right\rangle .  \notag
\end{eqnarray}%
In this state, $L_{\Vert }=S_{\Vert }=0$ and $L_{z}$ and $S_{z}$ are
constant. It is now the tilt angle of the pseudospin vector $\mathbf{O}$
that varies with bias. Orbital coherence begins when the bare energy of the
state $\left\vert K_{-},\pm ,0\right\rangle $ is equal to that of state $%
\left\vert K_{-},\pm ,1\right\rangle $ at $\nu =1,3.$ This occurs when%
\begin{eqnarray}
I_{1}^{\ast } &\rightarrow &O_{1}:\Delta _{B}^{\left( 12\right) }=\frac{%
\zeta _{1}}{\beta },  \label{orbit} \\
I_{3} &\rightarrow &O_{3}:\Delta _{B}^{\left( 17\right) }=\Delta
_{B}^{\left( 12\right) },
\end{eqnarray}%
and the critical bias does not depend on the magnetic field, Coulomb
interaction or on the value of the dielectric constant. We find $\zeta
_{1}/\beta =46$ meV i.e. $E=135$ mV/nm which is in the range of validity of
the two-band model.

The orbital phase survives until a transition to an incoherent phase occurs
at the critical bias%
\begin{eqnarray}
O_{1} &\rightarrow &I_{1}^{\ast \ast }:\Delta _{B}^{\left( 13\right) }=\frac{%
1}{\beta }\left( \zeta _{1}+\frac{1}{4}\Delta _{C}\right) , \\
O_{3} &\rightarrow &I_{3}^{\ast }:\Delta _{B}^{\left( 18\right) }=\Delta
_{B}^{\left( 13\right) }.
\end{eqnarray}%
Since $\zeta _{1}/\beta $ is the onset of the orbital phase, we see that the
range of existence of the orbital phase scales as $1/\sqrt{B}$. Part of this
range is outside the limit of validity of our model.

\begin{figure}[tbph]
\includegraphics[scale=0.8]{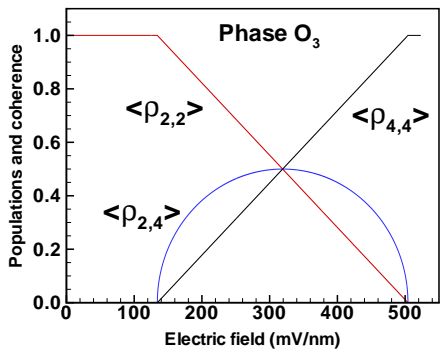}
\caption{(Color online) Variation of the populations interlayer coherence
with the transverse electric field in phase $O_{3}.$ }
\label{figpopO3}
\end{figure}

Fig. \ref{figpopO3} shows how the populations and coherence vary with the
transverse electric field in phase $O_{3}$. The same behavior is found in
phase $O_{1}.$ We remark that in previous work where the spin degree of
freedom is frozen\cite{CoteOrbital}, the orbital coherent phase occurs at $%
\nu =-1$ and $\nu =3$.

The orbital phase exists in a large range of bias. As we explained in Sec.
III.E, a finite orbital coherence implies a finite density of electric
dipoles in the plane of the layers. The orientation of these dipoles can be
controlled by an electric field in the plane of the layers\cite%
{Shizuyadipole}. On of us has studied in detail the interesting properties
of this state\cite{CoteOrbital,CoteOrbital2}. For example, the collective
mode associated with the orbital coherence is highly anisotropic. This mode
softens at a finite wave vector in the direction perpendicular to the
dipoles when the bias is increased. This suggests a transition to a
charge-density-wave state. In the Hartree-Fock approximation, it was found
that this transition is preempted by a transition to a crystal phase with
one electron per site and a Skyrmion-like pseudospin texture of the orbital
pseudospin at each crystal site. As the bias is increased, the crystal state
is followed by a helical state where the orbital pseudospin rotates along
one spatial direction. In both phases, the total electronic density is
modulated spatially but the local filling factor is not. By further
increasing the bias, the crystal state and then the uniform states are
recovered. The critical electric fields for the transition to the uniform
(UP), Skyrmion crystal (SKP) and helical phases (HP) are given by\cite%
{CoteOrbital2,note2}: 
\begin{equation}
\begin{tabular}{|l|l|}
\hline
$134$ $<\Delta _{B}<145$ mV/nm & UP \\ \hline
$145<\Delta _{B}<189$ mV/nm & SKP \\ \hline
$189$ $<\Delta _{B}<450\allowbreak $ mV/nm & HP \\ \hline
$450$ $<\Delta _{B}<494$ mV/nm & SKP \\ \hline
$494$ $<\Delta _{B}<$ $505$ mV/nm & UP \\ \hline
\end{tabular}
\label{nonphase}
\end{equation}%
The phase diagram in $O_{1}$ and $O_{3}$ is symmetric with respect to the
center of the helical phase.

We remark that this sequence of phase transitions is similar to that
observed in a thin film of the helical magnet Fe$_{0.5}$Co$_{0.5}$Si when a
perpendicular magnetic field is increased\cite{Jhhan}. It has been shown\cite%
{CoteOrbital2} that the Hamiltonian of the C2DEG in the orbital phase
contains a Dzyaloshinskii-Moriya (DM)\ interaction\cite{DM} that is
responsible for the rotation of the pseudospins. Its origin in bilayer
graphene is purely Coulombic while the DM\ interaction comes from spin-orbit
coupling in Fe$_{0.5}$Co$_{0.5}$Si.

We do not find any sign of instability in the collective mode dispersions
for the other phases in Fig. \ref{figdiagrammeglobal}. However, we remark
that phases with lower energy than those considered in this figure are
possible. In order to establish the phase diagram of the C2DEG, we choose a
set of possible ground states and compare their energies. This does not
ensure however that the true ground state is amongst the states that we have
chosen to compare! For the incoherent states, this is not a problem because
there is a finite number of states to compare. But for the nonuniform
states, the number of possible ground state is enormous.

\subsubsection{Spin-layer-coherent phases $SL_{\protect\nu }$}

The fourth type of phase has coherence between two states with the same
orbital index but different spin and layer indices. We use for these phases
the notation $SL_{\nu }.$ An example is phase $SL_{-1}$ where the ground
state is 
\begin{equation}
\left\vert \Psi _{SL_{-1}}\right\rangle =\prod\limits_{X}\left(
ac_{6,X}^{\dag }+bc_{1,X}^{\dag }\right) c_{7,X}^{\dag }c_{5,X}^{\dag
}\left\vert 0\right\rangle .
\end{equation}%
Because the coherence is now between two states with different spin \textit{%
and} layer indices, i.e. $\left\vert K_{+},-,0\right\rangle $ and $%
\left\vert K_{-},+,0\right\rangle $ in $\left\vert \Psi
_{SL_{-1}}\right\rangle $, we cannot describe the change with bias as the
tilting of one of the pseudospin $\mathbf{L},\mathbf{S}$ or $\mathbf{O}.$ In
fact this state has $L_{\Vert },S_{\Vert }=O_{\Vert }=0.$ Both $L_{z}$ and $%
S_{z}$ vary with bias however. This phase is characterized by the order
parameter $\left\langle \rho _{6,1}\right\rangle .$

The variation of $\left\langle \rho _{6,6}\right\rangle ,\left\langle \rho
_{1,1}\right\rangle $ and $\left\langle \rho _{6,1}\right\rangle $ with bias
in this phase is similar to that shown in Fig. \ref{figpopO3}. The critical
biases for the beginning ($\Delta _{b}\left( n\right) $) and end ($\Delta
_{e}\left( n\right) $) of the $SL_{\pm 1}$ phases are given by

\begin{eqnarray}
\Delta _{b}\left( n\right) &=&\frac{\frac{d}{\ell }\alpha +\Delta _{z}+X_{n}%
}{1-2\beta \delta _{n,1}}, \\
\Delta _{e}\left( n\right) &=&\frac{3\frac{d}{\ell }\alpha +\Delta _{z}-X_{n}%
}{1-2\beta \delta _{n,1}},
\end{eqnarray}%
where $X_{n}=X_{n,n,n,n}^{+,+}(0)-X_{n,n,n,n}^{+,-}(0).$ Thus, 
\begin{eqnarray}
\Delta ^{(5)} &\approx &\left( 2\frac{d}{\ell }-\sqrt{\frac{\pi }{8}}\left( 
\frac{d}{\ell }\right) ^{2}\right) \alpha +\Delta _{z}, \\
\Delta ^{(6)} &\approx &\left( 2\frac{d}{\ell }+\sqrt{\frac{\pi }{8}}\left( 
\frac{d}{\ell }\right) ^{2}\right) \alpha +\Delta _{z}, \\
\Delta ^{(10)} &\approx &\frac{\left( 2\frac{d}{\ell }-7\sqrt{\frac{\pi }{128%
}}\left( \frac{d}{\ell }\right) ^{2}\right) \alpha +\Delta _{z}}{1-2\beta },
\\
\Delta ^{(11)} &\approx &\frac{\left( 2\frac{d}{\ell }+7\sqrt{\frac{\pi }{128%
}}\left( \frac{d}{\ell }\right) ^{2}\right) \alpha +\Delta _{z}}{1-2\beta }.
\end{eqnarray}

These critical biases scale linearly with the magnetic field. The phase $%
SL_{\nu }$ is the ground state in a small range of bias of the order $\left(
d/\ell \right) ^{2}\alpha $ which is approximately $0.1$ meV for $\kappa =5.$

A sufficiently large bias is necessary for spin down states to cross the
spin up states and produces a $SL_{\nu }$ phase. The $SL_{0}$ is special
because it involves coherence in both $n=0$ and $n=1$. As for $L_{\pm 2}$,
the exact critical bias in this case has a complicated analytical expression
which we do not reproduce here.

\subsection{Spin polarization}

We indicate for each phase in Fig. \ref{figdiagrammeglobal} the spin
polarization $S_{z}$. The polarization is constant in all phases with the
exception of the phases $SL_{\nu }$ where it varies continuously between the
two numbers indicated. The biggest change in $S_{z}$ and $L_{z}$ occurs at
filling factor $\nu =0$ where the C2DEG makes a transition from a fully spin
polarized and layer unpolarized gas ($S_{z}=2\hslash ,L_{z}=0$) at small
bias to a spin unpolarized and layer polarized gas ($S_{z}=0,L_{z}=2$) at
large bias. For $\nu =\pm 1,$ the $SL_{\nu }$ phase interpolates between $%
S_{z}=3\hslash /2,L_{z}=1/2$ and $S_{z}=\hslash /2,L_{z}=3/2.$ The only
jumps in $S_{z}$ occur at the transitions $I_{\pm 2}\rightarrow I_{\pm
2}^{\ast }$ where the system goes from a spin polarized to a spin
unpolarized state.

\subsection{Transport gaps}

Another quantity that is accessible experimentally is the transport gap $%
\Delta $ which is defined by the difference in energy between the first
empty state and the last filled state of the Hartree-Fock Hamiltonian. It
was shown previously\cite{BarlasPRL1} that the gap at zero bias follow the
hierarchy $\Delta _{\nu =0}>\Delta _{\nu =\pm 2}>\Delta _{\nu =\pm 1,\pm 3}.$
This implies that the first plateau to appear when the magnetic field is
turned on has $\sigma _{xy}=0.$ At larger field, the $\sigma _{xy}=\pm 2$
plateaus appear and at still larger field, the $\sigma _{xy}=\pm 1,\pm 3$
plateaus. This is indeed what is seen experimentally\cite%
{Feldman,Martin,Zhao}.

\begin{figure}[tbph]
\includegraphics[scale=0.7]{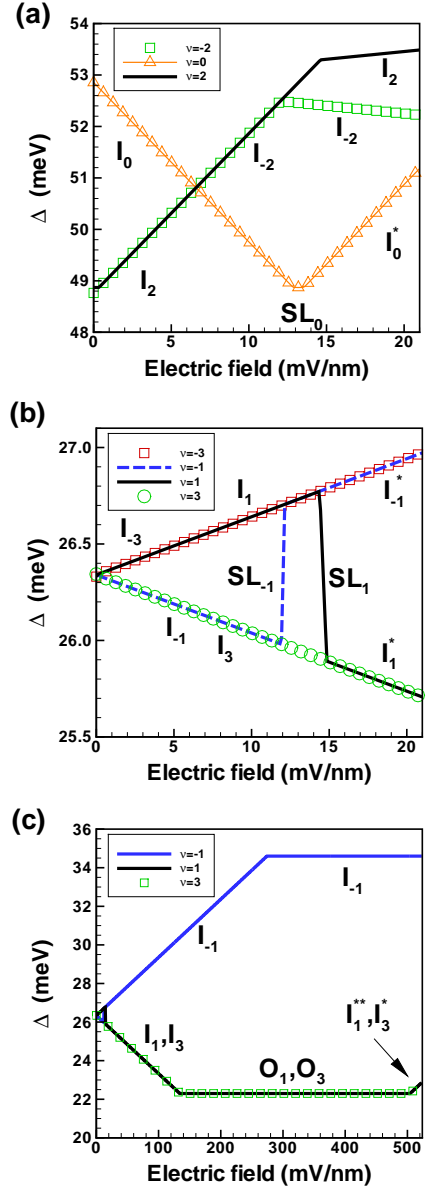}
\caption{(Color online) Variation of the transport gap with the
perpendicular electric field in the different phases of the C2DEG in Landau
level $N=0.$ The kink in the behavior of $I_{\pm 2}$ are due to level
crossings. A value of $\protect\kappa =5$ is assumed for the dielectric
constant and $B=10$ T. }
\label{figglobalgap}
\end{figure}

Fig. \ref{figglobalgap} shows the Hartree-Fock gaps as a function of the
transverse electric field for the different phases of the C2DEG. For this
figure, we have taken $B=10$ T and $\kappa =5.$ In some phases, one or more
level crossing occurs that change the behavior of the gap. This is clearly
visible for $I_{\pm 2}$ in Fig. \ref{figglobalgap} (a) and for $I_{-1}$ in
Fig. \ref{figglobalgap} (c).

With the exceptions of the phases where coherence occurs in two levels ($%
L_{\pm 2},SL_{0}$), it is possible to obtain a simple analytical expression
for the gap. We list these expressions below. When one or more level
crossings occur, we use the notation $I_{\nu }^{\left( 1\right) },I_{\nu
}^{\left( 2\right) },I_{\nu }^{\left( 3\right) },...$ to denote the
different behaviors of the gap and $\Delta _{B}^{\left( j\right) -\left(
j+1\right) }$ for the values at which the level crossings occur$.$

In the incoherent phases with $\nu =\pm 1,\pm 3:$

\begin{eqnarray}
I_{-3},I_{1},I_{-1}^{\ast \left( 1\right) } &:&\Delta =\beta \Delta
_{B}+\zeta _{1}+\frac{1}{2}\Delta _{C},  \label{gap1} \\
I_{3}^{\ast },I_{1}^{\ast \ast } &:&\Delta =\beta \Delta _{B}-\zeta _{1}+%
\frac{1}{4}\Delta _{C}.
\end{eqnarray}%
\begin{eqnarray}
I_{-1},I_{1}^{\ast },I_{3} &:&\Delta =-\beta \Delta _{B}+\zeta _{1}+\frac{1}{%
2}\Delta _{C},  \label{gap3} \\
I_{-1}^{\ast \left( 2\right) } &:&\Delta _{-1}^{\ast }=\Delta _{Z}+\frac{3}{4%
}\Delta _{C},
\end{eqnarray}%
where, for $\nu =-1:$%
\begin{equation}
\Delta _{B}^{\left( 1\right) -\left( 2\right) }=\frac{\Delta _{z}-\zeta _{1}+%
\frac{1}{4}\Delta _{C}}{\beta }.
\end{equation}%
For the incoherent phases with $\nu =-2$%
\begin{eqnarray}
I_{-2}^{\left( 1\right) } &:&\Delta =\left( 1-\beta \right) \Delta
_{B}-\zeta _{1}-2\frac{d}{\ell }\alpha +\frac{5}{4}\Delta _{C}, \\
I_{-2}^{\left( 2\right) } &:&\Delta =-\beta \Delta _{B}-\zeta _{1}+\Delta
_{Z}-2\frac{d}{\ell }\alpha +\frac{5}{4}\Delta _{C}, \\
I_{-2}^{\ast } &:&\Delta =\beta \Delta _{B}+\zeta _{1}-\Delta _{Z}+\frac{1}{2%
}\Delta _{C},
\end{eqnarray}%
with%
\begin{equation}
\Delta _{B}^{\left( 1\right) -\left( 2\right) }=\Delta _{Z}+2\frac{d}{\ell }%
\alpha ,
\end{equation}%
while for $\nu =2,$%
\begin{eqnarray}
I_{2}^{\left( 1\right) } &:&\Delta =\left( 1-\beta \right) \Delta _{B}-\zeta
_{1}-2\frac{d}{\ell }\alpha +\frac{5}{4}\Delta _{C}, \\
I_{2}^{\left( 2\right) } &:&\Delta =\beta \Delta _{B}+\Delta _{Z}-\zeta _{1}+%
\frac{5}{4}\Delta _{C}, \\
I_{2}^{\left( 3\right) } &:&\Delta =\Delta _{Z}+\frac{5}{4}\Delta _{C}, \\
I_{2}^{\left( 4\right) } &:&\Delta =-\beta \Delta _{B}+\Delta _{Z}+\zeta
_{1}+\frac{3}{2}\Delta _{C},
\end{eqnarray}%
with 
\begin{eqnarray}
\Delta _{B}^{\left( 1\right) -\left( 2\right) } &=&\frac{2\frac{d}{\ell }%
\alpha +\Delta _{z}}{1-2\beta }, \\
\Delta _{B}^{\left( 2\right) -\left( 3\right) } &=&\frac{\zeta _{1}}{\beta },
\\
\Delta _{B}^{\left( 3\right) -\left( 4\right) } &=&\frac{\zeta _{1}+\frac{1}{%
4}\Delta _{C}}{\beta }.
\end{eqnarray}%
For $\nu =0,$ we find: 
\begin{eqnarray}
I_{0} &:&\Delta =-\left( 1-\beta \right) \Delta _{B}-\zeta _{1}+\Delta _{Z}+%
\frac{5}{4}\Delta _{C}, \\
I_{0}^{\ast \left( 1\right) } &:&\Delta =\left( 1-\beta \right) \Delta
_{B}-\zeta _{1}-\Delta _{Z} \\
&&-4\frac{d}{\ell }\alpha +\frac{5}{4}\Delta _{C},  \notag \\
I_{0}^{\ast \left( 2\right) } &:&\Delta =\left( 1-2\beta \right) \Delta
_{B}-\Delta _{Z} \\
&&-4\frac{d}{\ell }\alpha +\frac{5}{4}\Delta _{C},  \notag
\end{eqnarray}%
with 
\begin{equation}
\Delta _{B}^{\left( 1\right) -\left( 2\right) }=\frac{\zeta _{1}}{\beta }.
\end{equation}%
The gap changes rapidly in the $SL_{\pm 1}$ phase while it is almost
independent of the bias in phase $SL_{0}$. Its value in the middle of the $%
SL_{\pm 1}$ phases is given approximately by%
\begin{eqnarray}
SL_{-1} &:&\Delta \approx \zeta _{1}+\frac{\Delta _{C}}{2}\left( 1-\frac{1}{4%
}\left( \frac{d}{\ell }\right) ^{2}\right) \\
&&-4\beta ^{2}\left( \frac{\Delta _{-1}}{\Delta _{C}}\right) ^{2}\frac{%
\Delta _{C}}{2},  \notag \\
SL_{1} &:&\Delta \approx \zeta _{1}+\frac{\Delta _{C}}{2}\left( 1+\frac{1}{8}%
\left( \frac{d}{\ell }\right) ^{2}\right) , \\
&&-4\beta ^{2}\left( \frac{\Delta _{1}}{\Delta _{C}}\right) ^{2}\frac{\Delta
_{C}}{2},  \notag
\end{eqnarray}%
where $\Delta _{\pm 1}$ are defined in Eqs. (\ref{gapm1},\ref{gapp1}). The
gaps are twice as big in phases $I_{0},I_{\pm 2}$ than in phases $I_{\pm
1},I_{\pm 3}$ and vary more rapidly with bias in the former than in the
latter. The presence of the $SL_{\nu }$ phase smoothens the jump of the gap
in the transition from $I_{\pm 1}$ to $I_{\pm 1}^{\ast }.$

The gap is independent of the bias in the orbital phases $O_{1}$ and $O_{3}$:%
\begin{equation}
O_{1},O_{3}:\Delta =\frac{\Delta _{C}}{2}.
\end{equation}

For the $L_{\nu }$ phases, the gaps at zero bias are given approximately by%
\begin{eqnarray}
L_{-3},L_{1} &:&\Delta \approx \zeta _{1}+\frac{1}{2}\left( 1-\frac{1}{4}%
\left( \frac{d}{\ell }\right) ^{2}\right) \Delta _{C}, \\
L_{-1},L_{3} &:&\Delta \approx \zeta _{1}+\frac{1}{2}\left( 1+\frac{1}{8}%
\left( \frac{d}{\ell }\right) ^{2}\right) \Delta _{C}, \\
L_{\pm 2}, &:&\Delta \approx -\zeta _{1}-2\left( \frac{d}{\ell }\right)
\alpha  \notag \\
&&+\frac{1}{4}\left( 5+\frac{23}{4}\left( \frac{d}{\ell }\right) ^{2}\right)
\Delta _{C}.
\end{eqnarray}%
The correction $\left( d/\ell \right) ^{2}$ is very small and the gaps at
zero bias for $L_{\pm 1}$ and $L_{\pm 3}$ are almost equal. (The difference
comes from the fact that the coherence is not in the same orbital in $%
L_{-3},L_{1}$ and $L_{-1},L_{3}$.) These gaps are not shown in Fig. \ref%
{figdiagrammeglobal} because the corresponding phases occur at very small
biases. The gap increases (decreases) with bias in phases $L_{-3},L_{1}$ ($%
L_{-1},L_{3}$)$.$ It is almost constant in $L_{\pm 2}$. The main
contribution to all gaps is the Coulomb exchange interaction.

With $\zeta _{1}=0,$ our gaps for phases $I_{2},I_{0},I_{0}^{\ast }$ agree
with those of Gorbar \textit{et al}.\cite{Gorbarnutot} if screening is
neglected in their calculation. For $I_{1},I_{1}^{\ast }$ and $I_{3},$
however, our exchange correction is $\Delta _{C}/2$ which is consistent with
Ref. \onlinecite{BarlasPRL1} while Gorbar \textit{et al}. have $3\Delta
_{C}/8.$

At $\Delta _{B}=0,$ we find for $I_{0}$ the gap $\Delta =62$ meV for $B=2$ T
and $\kappa =2$ while, with static screening, Gorbar et al. find $\approx 5$
meV. Similarly, for phase $I_{1},$ the HFA gives $\Delta =26$ meV while the
result with static screening is $\approx 2.5$ meV. Static screening leads to
a reduction of the gap by a factor of at least $10.$ Dynamical screening and
Landau mixing corrections, however, increase the gaps calculated with static
screening by a factor of two to three\cite{Gorbarconductivity}. As for the
behavior of the gap with bias. Fig. 5 of Gorbar \textit{et al}.\cite%
{Gorbarnutot} shows that, with screening, the gap of the phase $I_{3}$
increases with the electric field even when the correction $\beta \Delta
_{B} $ is neglected. The slope is approximately $0.1$ nm-C for $B=2$ T and $%
\kappa =2$. The (unscreened) HFA\ predicts a slope of $\beta d=0.006$ nm-C.
When both screening and $\beta \Delta _{B}$ are considered, the gaps for $%
I_{\pm 1},I_{\pm 3},I_{\pm 1}^{\ast }$ will probably increase with bias
(contrary to the behavior illustrated in our Fig. \ref{figglobalgap}(b)) but
the rapid change of the gap in the $SL_{\pm 1}$ will still be present.

The energy gaps obtained from local compressibility measurements on
suspended bilayer graphene by Martin \textit{et al}.\cite{Martin} are of
size $\Delta _{\nu =0}\approx 1.7B$ $\left[ \text{T}\right] $ meV, $\Delta
_{\nu =\pm 2}\approx 1.2B$ $\left[ \text{T}\right] $ meV and $\Delta _{\nu
=\pm 1}\approx 0.1B$ $\left[ \text{T}\right] $ meV (with less data points in
this case). A more recent transport experiment by Velasco \textit{et al}.%
\cite{Velasco} on suspended bilayer graphene with a higher mobility reports
a larger gap $\Delta _{\nu =0}\approx 5.5B$ $\left[ \text{T}\right] $ meV.
The measured gaps scale linearly with the magnetic field contrary to the HFA
prediction. In fact, Gorbar \textit{et al}..\cite{Gorbarnutot} have shown
that a linear scaling is obtained if static screening is considered. (One
set of experiments at higher magnetic field reported gaps that scaled as $%
\sqrt{B}$ however\cite{Zhao}.)

Experiments\cite{Weitz} show that $\sigma _{xy}$ ceases to be quantized at $%
\nu =0,1$ in the region corresponding to the $SL_{\nu }$ phase and at $\nu
=2 $ and $\nu =3$ in the region around zero bias. A possible explanation is
that the conductance quantization is broken by disorder in the regions
corresponding to a minimum of the gap\cite{Gorbarnutot}. However, that
argument does not seem to work at $\nu =1$ where the screened HFA\ gap
increases continuously with the bias and is also not compatible with our
unscreened result.

\subsection{Collective modes and optical absorption}

\begin{figure*}[tbph]
\includegraphics[scale=0.8]{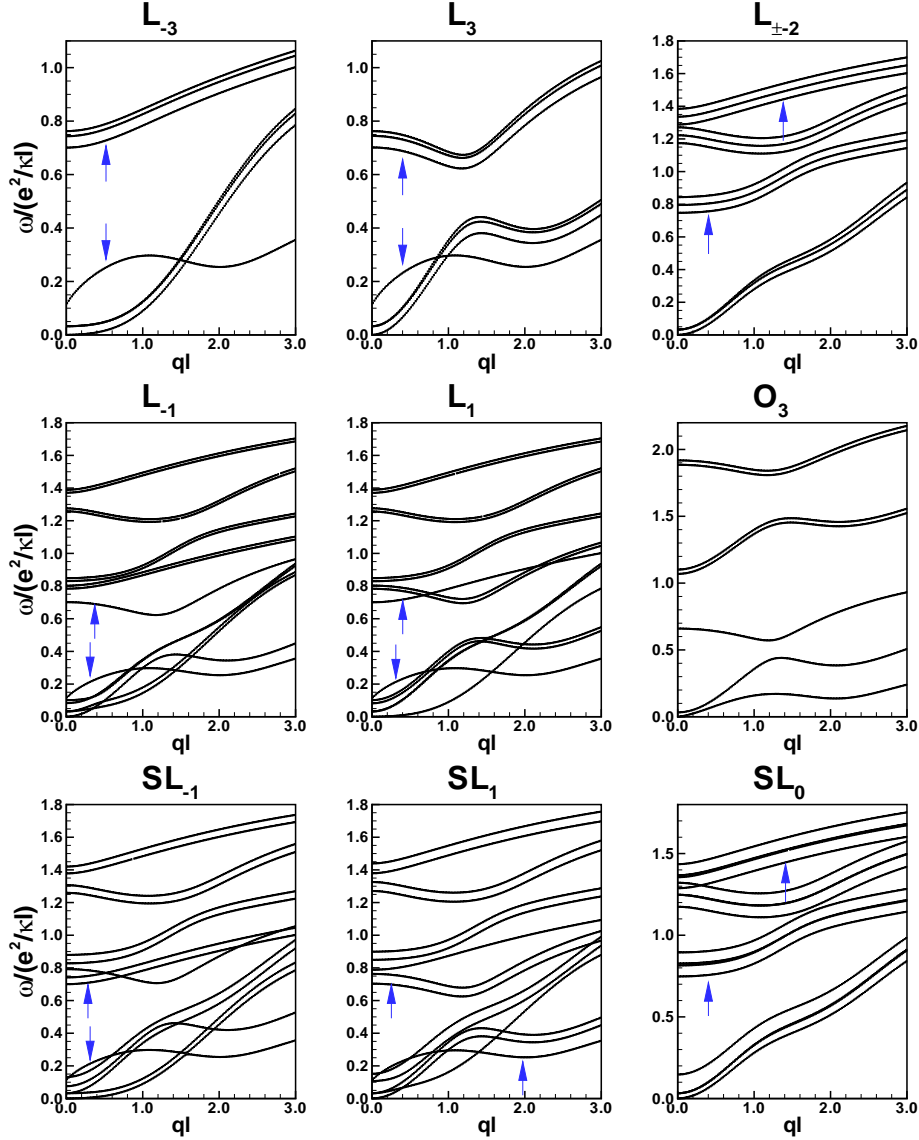}
\caption{(Color online) Dispersion of the collective modes in the coherent
phases. The (blue) arrows point to the modes which are active in optical
absorption experiments. All modes are evaluated at $B=10$ T with $\protect%
\kappa =5.$ The bias in units of $e^{2}/\protect\kappa \ell =35.6$ meV
(corresponding to a frequency $\protect\nu =e^{2}/h\protect\kappa \ell
=8.6\times 10^{12}$ Hz) is respectively: $\Delta _{B}=0.0005$ for $%
L_{-3},L_{1};$ $\Delta _{B}=0.001$ for $L_{3},L_{-1};\Delta _{B}=0.002$ for $%
L_{-2};\Delta _{B}=1.26$ for $O_{3};$ $\Delta _{B}=1.115$ for $%
SL_{-1};\Delta _{B}=0.14$ for $SL_{1}$ and $\Delta _{B}=0.126$ for $SL_{0}.$
In $SL_{0}$ the middle line in each group of three dispersive curves
contains two modes which are very close in energy.}
\label{figmodes}
\end{figure*}

\begin{figure*}[tbph]
\includegraphics[scale=0.9]{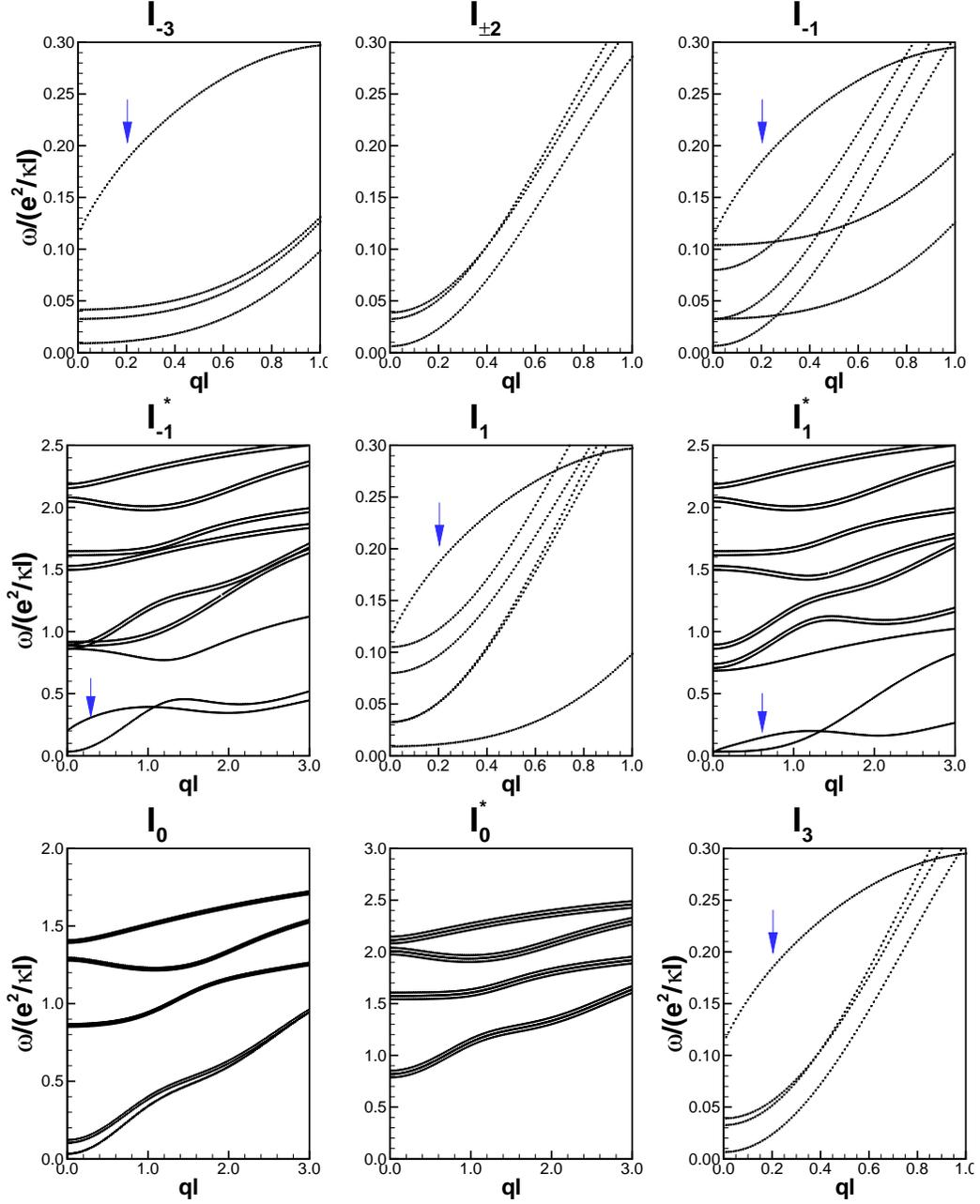}
\caption{(Color online) Dispersion of the collective modes in the incoherent
phases for $B=10$ T and $\protect\kappa =5.$ The bias is in units of $e^{2}/%
\protect\kappa \ell =35.6$ meV corresponding to a frequency $\protect\nu %
=e^{2}/h\protect\kappa \ell =8.6\times 10^{12}$ Hz. The (blue) arrows point
to the modes which are active in optical absorption experiments. The bias in
units of $e^{2}/\protect\kappa \ell $ is respectively: $\Delta _{B}=0.01$
for $I_{0},I_{\pm 1},I_{\pm 2},I_{\pm 3}$ and $\Delta _{B}=1.0$ for $%
I_{0}^{\ast },I_{\pm 1}^{\ast }.$ For $I_{\pm 3}$,$I_{\pm 2},I_{\pm 1}$ only
the first four, three, and six modes respectively are shown. For $I_{0}$,
each of the four branches contains four modes which are close in energy. In $%
I_{0}^{\ast }$ the middle line in each group of three curves contains two
modes.}
\label{figmodeio}
\end{figure*}

Each phase of the C2DEG is characterized by a set of collective excitations.
The number of dispersive modes when $m$ levels are filled is $m\left(
8-m\right) $. We have calculated the dispersion relation of these modes
using the GRPA\ described in Sec. IIII. In the limit $\mathbf{q}\rightarrow
\infty $, the vertex corrections vanish and the response function $\chi
\rightarrow \chi ^{0}$ where $\chi ^{0}$ is the response function evaluated
in the HFA. Thus, in this limit, the collective mode frequencies must
correspond to transitions between a filled and an empty eigenstate of the
Hartree-Fock Hamiltonian. At finite value of $\mathbf{q},$ some modes mix
together and it becomes difficult to identify their character (layer,
orbital, spin transitions, etc.). Our numerical results are summarized in
Fig. \ref{figmodes} for the coherent phases and in Fig. \ref{figmodeio} for
the incoherent phases.

\subsubsection{Goldstone modes}

The coherent phases sustain one gapless (Goldstone) mode. The number of
Goldstone modes is indicated for each phase in Fig. \ref{figdiagrammeglobal}%
. For example, in phase $L_{-3},$ this mode is due to the fact that the
layer pseudospin $\mathbf{L}$ can rotate freely around the $\widehat{\mathbf{%
z}}$ axis. The same situation occurs for the coherent phase $O_{\nu }$ where
again the orbital pseudospin $\mathbf{O}$ can rotate freely around the $%
\widehat{\mathbf{z}}$ axis. Phases $L_{\pm 2}$ support coherence in both $%
n=0 $ and $n=1$ orbitals and we can define a layer pseudospin $\mathbf{L}%
_{0} $ for $n=0$ and $\mathbf{L}_{1}$ for $n=1.$ The Goldstone mode in this
case correspond to an in-phase rotation of both pseudospins. Alternatively,
we can see this mode as a fluctuation of the relative phase of the two order
parameters $\left\langle \rho _{1,5}\right\rangle $ and $\left\langle \rho
_{3,7}\right\rangle $ in $L_{-2}$ or $\left\langle \rho _{2,6}\right\rangle $
and $\left\langle \rho _{4,8}\right\rangle $ in $L_{2}$.

The dispersion off all gapless modes (with the exception of the gapless mode
in the orbital phase) is linear in wave vector at very small wave vector
i.e. for $q\ell \lesssim d/\ell .$ Phases $L_{\pm 2}$ have the same
collective mode spectrum. The dispersion of the Goldstone mode in phases
where the coherence occurs in the orbital $n=1$ (i.e. $L_{-1},L_{3},SL_{1}$)
has a roton minimum while there is none if the coherence occurs in $n=0$
(i.e. $L_{-3},L_{1},SL_{-1}$). This is due to the particular form factor for 
$n=1$ involved in the Coulomb matrix elements (see Eqs. (\ref{form1}-\ref%
{form4})). Phases $L_{-2},L_{2},SL_{0}$ contain coherence in both $n=0$ and $%
n=1$ and a small shoulder appears in the dispersion.

For phase $L_{\nu },$ the Goldstone mode is the famous layer-pseudospin-wave
mode which has been extensively studied in semiconductor bilayer\cite%
{Layerpseudospin} at filling factor $\nu =1$ and detected experimentally\cite%
{Spielman}. In semiconductor bilayer, this mode becomes soft at a finite
wave vector as the separation between the layers is increased (around $%
d/\ell \approx 1$). In bilayer graphene $d/\ell <<1$ and the layer-coherent
phases are stable. The only instability in the collective modes is seen in
the orbital phases $O_{1}$ and $O_{3}.$

For $\zeta _{1}=0,$ the Goldstone mode of $L_{-3}($and $L_{1}$) has a
quadratic dispersion at zero bias and becomes unstable\cite{BarlasPRL2} at
finite bias. (We have checked that these conclusions remain valid if $\zeta
_{1}$ is finite but small.) A consequence of this instability is that the
C2DEG is expected to go from a smectic (non-homogeneous)\ phase at low
temperature to an isotropic phase at higher temperature. The smectic phase
would lead to anisotropic electrical transport.

The dispersion of the gapless orbital pseudospin-wave mode was studied in
detail for the uniform phase\cite{CoteOrbital} as well as for the crystal
and helical phases\cite{CoteOrbital2}. In the uniform phase, it has a
strongly anisotropic dispersion: linear in the direction of the orbital
pseudospins and quadratic in the other directions i.e.%
\begin{eqnarray}
\omega \left( \mathbf{q}\right) &=&\sqrt{2\left( \beta \Delta _{B}-\zeta
_{1}\right) q\ell }\left\vert \sin \left( \theta _{\mathbf{q}}\right)
\right\vert ,  \label{dis1} \\
\omega \left( \mathbf{q}\right) &=&\frac{1}{4}\sqrt{\sqrt{2\pi }\left( \beta
\Delta _{B}-\zeta _{1}\right) }q\ell ,(\theta _{\mathbf{q}}=0,\pi ),
\label{dis2}
\end{eqnarray}%
where $\theta _{\mathbf{q}}$ is the angle between the wave vector and the $x$
axis. The Goldstone mode softens at a finite wave vector $q\ell \approx 2$
in the direction perpendicular to the orbital pseudospins at a bias $\Delta
_{B}=58.8$ meV ($E=173$ mV/nm). This suggests a transition to a
charge-density-wave state. As we explained above, this transition is
preempted by a transition to a crystal phase at $\Delta _{B}=47.7$ meV.

\subsubsection{Spin-wave modes}

The number of spin-wave modes gapped at $\Delta _{Z}$ is indicated for each
phase in Fig. \ref{figdiagrammeglobal}.

All coherent phases with the exception of $O_{1},O_{3}$ have two modes
gapped at the Zeeman energy $\Delta _{Z}$ at $\mathbf{q}=0.$ Their
degeneracy is lifted at finite wave vector. Because of the occupation of the
levels in $O_{1}$ and $O_{3},$ only one intralayer spin-flip transition is
possible in theses phases.

The incoherent phases can have $0,1$ or $2$ modes gapped at $\Delta _{Z}.$
To be gapped at $\Delta _{Z},$ these modes must involve transitions within
the same valley and orbitals. From Fig. \ref{figdiagrammeglobal}, it is easy
to see that no intralayer and intraorbital spin-flip transition is possible
for $I_{0}^{\ast }$ and $I_{\pm 2}^{\ast }$ and that the occupation of the
levels permit only one such mode in $I_{\pm 3},I_{\pm 1}^{\ast }.$ In $%
I_{\pm 2},$ two transitions seem possible but they are degenerate and the
coupling between them leaves one mode gapped at $\Delta _{Z}$ and the second
mode has its frequency renormalized. The same mechanism operates in phases $%
I_{\pm 1},I_{0}$ resulting in two modes gapped at $\Delta _{Z}.$

\subsubsection{Optical absorption}

In the absence of Coulomb interaction, the dynamical conductivity has
intra-octet peaks at the bare gap energy $\Delta _{\xi }^{0}=E_{\xi
,1}^{0}-E_{\xi ,0}^{0}=\zeta _{1}+\xi \beta \Delta _{B}$ in addition to the
inter-Landau-level peaks which do not appear in our calculation. Fig. \ref%
{figabso} shows the absorption in different phases when Coulomb interaction
is considered. The number of absorption peaks is also indicated for each
phase in Fig. \ref{figdiagrammeglobal} and, in Figs. \ref{figmodes},\ref%
{figmodeio}, we identify by (blue) arrows the modes that lead to optical
absorption.

In the incoherent phases, optical absorption is possible only for
transitions that occur between states with the same valley and spin indices
but different orbital indices. The incoherent phases have one or zero
absorption peak. The latter case occurs when both levels $n=0,1$ with the
same layer and spin indices are filled (phases $I_{\pm 2},I_{0},I_{0}^{\ast
} $).

When level $n=0$ is filled, the absorption, as shown in Sec. III.g is
exactly at $\omega =\left( \zeta _{1}+\xi \beta \Delta _{B}\right) /\hslash
. $ When level $n=1$ is filled, the absorption is at $\omega =\left( \zeta
_{1}-\beta \Delta _{B}+\Delta _{C}/4\right) /\hslash $ i.e. affected by
exchange corrections. The former case applies to most of the incoherent
phases in our phase diagram. The latter case applies to phases $I_{1}^{\ast
\ast },I_{2}^{\ast }$ and $I_{3}^{\ast }$ which occur at very high bias and
are outside the limits of validity of our two-band model. In those phases,
level $n=1$ is filled because it is below $n=0$ in energy. In phases $I_{\pm
2}^{\ast }$, the two allowed transitions have the same energy and the
intensity of the absorption peak is doubled.

\begin{figure}[tbph]
\includegraphics[scale=0.8]{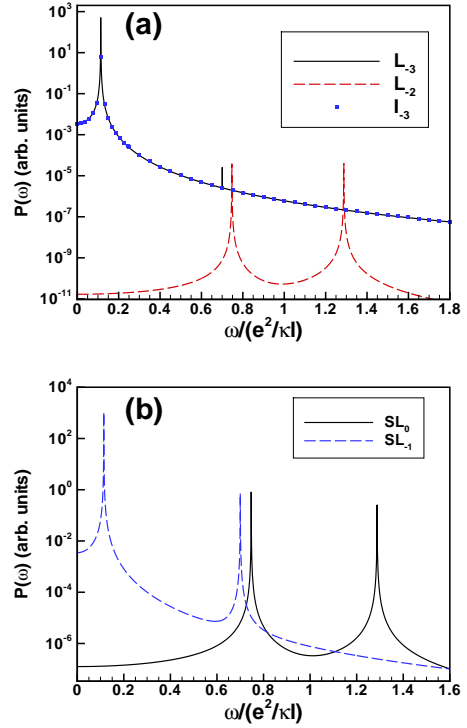}
\caption{(Color online) Electromagnetic absorption for $B=10$ T and $\protect%
\kappa =5$ in phases (a) $L_{-3}$ ($\Delta _{B}=0.018$ meV)$,$ $I_{3}$ ($%
\Delta _{B}=0.36$ meV) and $L_{-2}$ ($\Delta _{B}=0.07$ meV); (b) $SL_{0}$ ($%
\Delta _{B}=4.48$ meV) and $SL_{-1}$ ($\Delta _{B}=4.09$ meV)$.$ }
\label{figabso}
\end{figure}

Fig. \ref{figabso} (a) shows the absorption as a function of frequency in
phase $I_{-3}$ at finite bias. The absorption is concentrated in one strong
peak at the frequency 
\begin{eqnarray}
\nu &=&\left( \zeta _{1}+\xi \beta \Delta _{B}\right) /h \\
&=&\left( 9.\,\allowbreak 8+7.6\xi \Delta _{B}\left[ \text{in }\frac{e^{2}}{%
\kappa \ell }\right] \right) \times 10^{11}\text{ Hz}  \notag
\end{eqnarray}%
at $B=10$ T and $\kappa =5.$ The absorption frequency varies widely with bias%
$.$ For example, at the onset of the transition from $I_{1}^{\ast }$ to $%
O_{1},$ the frequency $\nu \rightarrow 0$ (see Eq. (\ref{orbit})) while $\nu
\approx 9.8\times 10^{11}$ Hz at the onset of the $I_{1}$ phase. By
contrast, if the absorption occurs in layer $K_{+},$ the minimum frequency
is $\nu =9.8\times 10^{11}$ Hz since the absorption frequency must increase
with bias in this case.

In the $O_{1},O_{3}$ phases, the Goldstone mode has orbital character (i.e.
electric dipole fluctuations) but does not lead to absorption at \textit{%
finite} frequency.

Interestingly, the coherent phases $L_{\nu }$ and $SL_{\nu }$ show two
absorption peaks (see Fig. \ref{figabso}). The second peak in $L_{\pm
1},L_{\pm 3}$ is extremely weak however and is absent at zero bias. The two
weak peaks in $L_{\pm 2}$ are also extremely weak and disappear at zero
bias. In $L_{-3},L_{1},$ the first absorption peak at zero bias is exactly
at $\hslash \omega =\zeta _{1}$ while for $L_{-1},L_{3},$ the frequency is
at $\hslash \omega =\zeta _{1}+\left(
X_{0,0,0,0}^{+,+}-X_{0,0,0,0}^{+,-}-X_{1,1,1,1}^{+,+}+X_{1,1,1,1}^{+,-}%
\right) /2$ which is slightly shifted from $\hslash \omega =\zeta _{1}.$

The layer eigenstates of the Hartree-Fock Hamiltonian at zero bias are the
symmetric ($S$) and antisymmetric ($AS$)\ combinations of $K_{+}$ and $%
K_{-}. $ Using Eqs. (\ref{chizero}) and (\ref{grpa2}), we can show from the
GRPA\ equations that the response functions that enter in Eq. (\ref{abso})
for the absorption in the absence of orbital coherence and at zero bias are
of the form $\chi _{0,1,1,0}^{\left( a_{\xi },\sigma _{1}\right) ,\left(
a_{\xi },\sigma _{1}\right) ,\left( a_{\xi },\sigma _{1}\right) ,\left(
a_{\xi },\sigma _{1}\right) }\left( \omega \right) $ or $\chi
_{1,0,01}^{\left( a_{\xi },\sigma _{1}\right) ,\left( a_{\xi },\sigma
_{1}\right) ,\left( a_{\xi },\sigma _{1}\right) ,\left( a_{\xi },\sigma
_{1}\right) }\left( \omega \right) $ where $a_{\xi }$ now stands for the $S$
or $AS$ layer combinations. That is, the layer combination is conserved in
the absorption at zero bias and in the absence of orbital coherence. It
follows that, in phase $L_{-3}$, the state $\left\vert S,+,0\right\rangle $
is filled and only the transition $\left\vert S,+,0\right\rangle \rightarrow
\left\vert S,+,1\right\rangle $ is optically active. For $L_{-2},$ no
transition conserving the valley index ($S$ or $AS$) is possible and for $%
L_{-1}$ the only allowed transition is between $\left\vert
AS,+,0\right\rangle \rightarrow \left\vert AS,+,1\right\rangle $ since
levels $\left\vert S,+,0\right\rangle ,\left\vert S,+,1\right\rangle
,\left\vert AS,+,0\right\rangle $ are filled. The same argument applies to
the spin down states.

At finite but small bias, the $S$ and $AS$ layer combinations are replaced
by bonding and anti-bonding combinations i.e. $\left\vert B,+,0\right\rangle
=a\left\vert K_{-},+,0\right\rangle +b\left\vert K_{+},+,0\right\rangle $
and $\left\vert AB,+,0\right\rangle =-b\left\vert K_{-},+,0\right\rangle
+a\left\vert K_{+},+,0\right\rangle $ for example where $a$ and $b$ depend
on the bias and on the orbital index $n$. The strong peak in the absorption
corresponds to the transition $\left\vert B,+,0\right\rangle \rightarrow
\left\vert B,+,1\right\rangle $ for $L_{-3}$ and to $\left\vert
AB,+,0\right\rangle \rightarrow \left\vert AB,+,1\right\rangle $ for $%
L_{-1}. $ By contrast to the zero bias case, the absorption given by Eq. (%
\ref{absoabso}) contains response functions which are not just of the form $%
\chi _{0,1,1,0}^{\left( a_{\xi },\sigma _{1}\right) ,\left( a_{\xi },\sigma
_{1}\right) ,\left( a_{\xi },\sigma _{1}\right) ,\left( a_{\xi },\sigma
_{1}\right) }\left( \omega \right) $ with $a_{\xi }=B,AB$ so that
transitions that do not conserve the layer combination $B,AB$ are weakly
optically active\cite{Note3}. A second peak appears in the absorption
spectrum which corresponds to the transition $\left\vert B,+,0\right\rangle
\rightarrow \left\vert AB,+,1\right\rangle $ for $L_{-3}$ and $L_{-1}.$ The
two weak peaks in $L_{-2}$ come from the transitions $\left\vert
B,+,0\right\rangle \rightarrow \left\vert AB,+,1\right\rangle $ and $%
\left\vert B,+,1\right\rangle \rightarrow \left\vert AB,+,0\right\rangle .$

Fig. \ref{figabso} (b) shows the absorption in phases $SL_{0}$ and $SL_{-1.}$
In $SL_{-1},$ we find by analyzing the eigenvectors of the modes involved in
the optical absorption that the strong peak corresponds to the transition $%
\left\vert B,\eta ,0\right\rangle \rightarrow $ $\left\vert B,\eta
,1\right\rangle $ and the weak peak to the transition $\left\vert B,\eta
,0\right\rangle \rightarrow $ $\left\vert AB,\eta ,1\right\rangle $ where $%
\left\vert B,\eta ,0\right\rangle =a\left\vert K_{-},+,0\right\rangle
+b\left\vert K_{+},-,0\right\rangle $ and $\left\vert AB,\eta
,0\right\rangle =-b\left\vert K_{-},+,0\right\rangle +a\left\vert
K_{+},-,0\right\rangle $ and $a,b$ depend on the bias and on the orbital
index $n$. The layer combination here are between two states with opposite
spin orientations.

The electromagnetic absorption in the crystal and helical phases are much
more complex and was discussed previously\cite{CoteOrbital2}. In the helical
phase, for example, the absorption depends on the orientation of the
polarization of the electromagnetic wave in the $x-y$ plane.

The collective mode dispersions computed here for the intra-Landau level ($%
N=0$) transitions should be combined with the dispersion of the inter-Landau
level magnetoexcitons computed by Sari and T\"{o}ke\cite{Sari} and by Shizuya%
\cite{Shizuya5} to get a complete picture of the absorption for the C2DEG in
bilayer graphene.

\section{EFFECT OF AN IN-PLANE ELECTRIC FIELD}

In this section, we consider the effect of adding a uniform in-plane
electric field to the Hartree-Fock Hamiltonian of Eq. (\ref{HHF2}). The
coupling between the C2DEG and this external electric field is given by

\begin{equation}
H_{E}=-\mathbf{d}\cdot \mathbf{E}_{\Vert },
\end{equation}%
where $\mathbf{d}$ is the total dipole moment of the electron gas given by
Eq. (\ref{dipole}).

The main effect of $\mathbf{E}_{\Vert }$ is to induce orbital coherence. We
consider here the case of $\nu =-1$ but a similar effect occur at other
filling factors and will be discussed elsewhere\cite{multiferro}. We set $%
\mathbf{E}_{\Vert }=-E_{0}\widehat{\mathbf{x}}.$ In the ground state, the
electric dipoles are aligned with $\mathbf{E}_{\Vert }.$

\begin{figure}[tbph]
\includegraphics[scale=0.8]{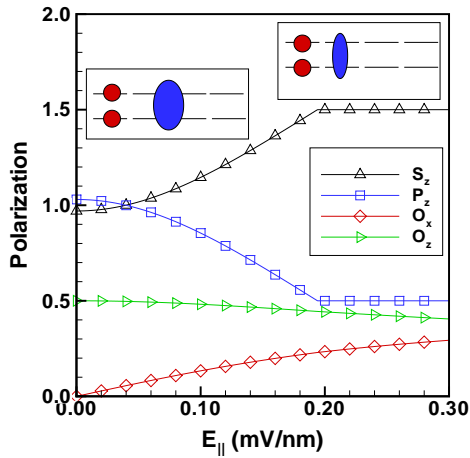}
\caption{(Color online) Variation of the spin and layer polarization and of
the orbital coherence with an applied in-plane electric field in phase $%
SOL_{-1}.$ The bias $\Delta _{B}$ has been taken near the middle of the $%
SL_{-1}$ phase where $\left\langle S_{z}\right\rangle =\left\langle
P_{z}\right\rangle \approx 1.$ Parameters are $B=10$ T and $\protect\kappa %
=5.$}
\label{multiferro}
\end{figure}

With finite $\mathbf{E}_{\Vert },$ the $SL_{-1}$ state is replaced by a
state with spin, orbital, and layer coherences i.e. $SOL_{-1}$. This is
represented by the inset in the top-left corner of Fig. \ref{multiferro}.
The wave function of the ground state becomes 
\begin{eqnarray}
\left\vert \Psi _{SOL_{-1}}\right\rangle &\rightarrow
&\prod\limits_{X}\left( ac_{6,X}^{\dag }+bc_{1,X}^{\dag }+cc_{3,X}^{\dag
}+dc_{8,X}^{\dag }\right) \\
&&\times c_{7,X}^{\dag }c_{5,X}^{\dag }\left\vert 0\right\rangle  \notag
\end{eqnarray}%
with 
\begin{equation}
\left\vert a\right\vert ^{2}+\left\vert b\right\vert ^{2}+\left\vert
c\right\vert ^{2}+\left\vert d\right\vert ^{2}=1.  \label{sum}
\end{equation}%
We show in Fig. \ref{multiferro} how the different polarizations $%
L_{z},P_{z},O_{z}$ and $O_{x}$ change with bias in $SL_{-1}$ when $\mathbf{E}%
_{\Vert }$ is increased. The other polarizations are zero. The spin(layer)
polarization $S_{z}\left( L_{z}\right) $ increases(decreases) with with $%
\mathbf{E}_{\Vert }$ until $E_{\Vert }^{c}\gtrsim 0.2$ mV/nm where it
remains constant. The orbital coherence has not yet saturated at $E_{\Vert
}^{c}$. The ground state above this critical electric field is represented
in the inset at the top-right corner of Fig. \ref{multiferro}. It is
interesting that the \textit{spin} polarization can be varied in this phase
by an external \textit{electric} field.

The state $\left\vert \Psi _{SOL_{-1}}\right\rangle $ has one gapless
Goldstone mode and its dispersion is anisotropic in wave-vector space (as is
the dispersion of the other modes). By contrast, if we apply $\mathbf{E}%
_{\Vert }$ to a phase $O_{\nu }$, the U(1) symmetry of the dipoles in the $%
x-y$ plane is broken. The orbital pseudospin $\mathbf{O}$ is then forced to
align with $\mathbf{E}_{\Vert }$ and the Goldstone mode is gapped\cite%
{Shizuyadipole}.

\begin{figure}[tbph]
\includegraphics[scale=0.8]{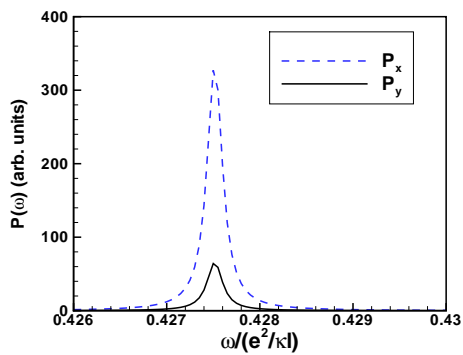}
\caption{(Color online) Electromagnetic absorption in the $SOL_{-1}$ phase
for two different polarizations of the electromagnetic field.}
\label{absoSLO}
\end{figure}

The absorption in $\left\vert \Psi _{SOL_{-1}}\right\rangle $ shows two
peaks as in $\left\vert \Psi _{SL_{-1}}\right\rangle .$ The first,
low-energy, peak is shown in Fig. \ref{absoSLO} for two different
orientations of the electromagnetic wave polarization. Clearly, the
absorption is anisotropic in phase $SOL_{-1}$ by contrast to all the other
uniform states that we studied before.

\section{CONCLUSION}

In this work, we have derived the phase diagram of the C2DEG in a
Bernal-stacked graphene bilayer. For the non-interacting Hamiltonian, we
used a tight-binding model with the hopping terms $\gamma _{0},\gamma
_{1},\gamma _{4}$ and $\delta $ and introduced a potential bias between the
two layers given by $\Delta _{B}.$ The Coulomb interaction was treated in
the Hartree-Fock approximation. To reduce the complexity of the problem, we
used an effective two-band model which described the low-energy behavior of
the C2DEG and is valid for $\Delta _{B}<<\gamma _{1}.$ We also restricted
the Hilbert space to the $N=0$ Landau level only and worked at zero
temperature. Our method allows us to include both coherent and incoherent
phases in the phase diagram. Indeed, we found phases with layer coherence at
small bias, spin and layer coherence at intermediate bias and orbital
coherence at large bias. The application of a parallel electric field, as we
showed, can also lead to a state with orbital, layer and spin coherence.

We have included in our analysis the hopping parameter $\gamma _{4}$ which
is often neglected in theoretical calculations. In our calculations, we find
that the phase diagram in sensitive to the precise value of this parameter.
If the value of this parameter is modified, the phases that we have
discussed are still present in the phase diagram but they occur at different
bias. Moreover, other phases may appear. For example, with $\gamma _{4}=0,$
the orbital-coherent phase becomes possible at $\nu =-1$ and a new phase
with valley and orbital coherence appears as discussed in Ref. %
\onlinecite{CoteOrbital}.

We have written down the ground-state wave function for each phase in the
global phase diagram of the C2DEG. We have also calculated for each phase
the transport gap, the spin polarization, the collective mode dispersions
and the electromagnetic absorption spectrum. The change in these properties
from one phase to another should facilitate their experimental
identification. Strictly speaking, however, our results are only valid
within the limits of validity of the approximations listed above. In
particular, we have neglected screening corrections which are known to
reduce significantly the transport gaps. These corrections were considered
for the incoherent phases in Ref. \onlinecite{Gorbarnutot}. In principle,
these screening corrections should be smaller at larger magnetic field. The
stability of the different phases that we found should also be studied by
considering quantum and thermal fluctuations as well as disorder effects.

More subtle corrections specific to graphene have also been considered by
Shizuya\cite{Shizuyacyclotron,Shizuyadipole,Shizuyamodes01}. The quantum
fluctuations of the Dirac sea (the filled Landau levels from the valence
band) have been shown to be sizable and to lead to corrections of the energy
of the octet of states in $N=0$\cite{Shizuya2012}. According to Shizuya, the
orbital degeneracy of the zero-energy levels is lifted by Coulombic vacuum
fluctuations, leading to an appreciable shift and splitting of the $n=0$ and 
$n=1$ levels and to a negative capacitance effect that blocks the rotation
of the valley pseudospins. The negative capacitance effect appears when the
full four bands of the tight-binding model are considered.

A more complete calculation would include all these effects and allow a more
direct comparison with the experimental results.

\begin{acknowledgments}
R. C\^{o}t\'{e} was supported by a grant from the Natural Sciences and
Engineering Research Council of Canada (NSERC). Computer time was provided
by Calcul Qu\'{e}bec and Compute Canada.
\end{acknowledgments}


\begin{thebibliography}{99}
\bibitem{note2} With $\zeta _{1}\neq 0,$ the beginning of the orbital phase
is at $\Delta _{B}=\zeta _{1}/\beta $ and not at $\Delta _{B,onset}+\zeta
_{1}/\beta $ as erroneously mentionned on page 8 of Ref. %
\onlinecite{CoteOrbital2}.
\end{thebibliography}
\end{document}